\documentclass[a4paper,fleqn]{cas-dc}

\usepackage[numbers]{natbib}

\usepackage{graphicx}
\usepackage{float}
\usepackage{subfigure}
\usepackage{wrapfig}

\def\tsc#1{\csdef{#1}{\textsc{\lowercase{#1}}\xspace}}
\tsc{WGM}
\tsc{QE}
\tsc{EP}
\tsc{PMS}
\tsc{BEC}
\tsc{DE}

\usepackage{caption}
\usepackage[dvipsnames]{xcolor}  
\usepackage{listings}  

\begin{document}

\shorttitle{Chenggang Shan et~al.~KubeAdaptor: A Docking Framework for Workflow Containerization on Kubernetes}    

\shortauthors{Chenggang Shan et~al.}  

\title [mode = title]{KubeAdaptor: A Docking Framework for Workflow Containerization on Kubernetes}  



%

\author[1,2]{Chenggang Shan}


\ead{uzz_scg@163.com}
\address[1]{School of Automation, Beijing Institute of Technology, Beijing 100081, China}


\author[1,3]{Guan Wang}
\ead{netspecters@126.com}

\author[1]{Yuanqing Xia}


\ead{xia_yuanqing@bit.edu.cn}




\author[1]{Yufeng Zhan}
\ead{yu-feng.zhan@bit.edu.cn}
\address[2]{School of Artificial Intelligence, Zaozhuang University, Zaozhuang 277100, China}

\author[1]{Jinhui Zhang}
\cormark[1]
\ead{zhangjinh@bit.edu.cn}
\address[3]{School of Information Science and Engineering, Zaozhuang University, Zaozhuang 277100, China}

\cortext[cor1]{Corresponding author}





\begin{abstract}
As Kubernetes becomes the infrastructure of the cloud-native era, the integration of workflow systems with 
Kubernetes is gaining more and more popularity. 
To our knowledge, workflow systems employ scheduling algorithms that optimize task execution order of workflow 
to improve performance and execution efficiency. However, due to its inherent scheduling mechanism, 
Kubernetes does not execute containerized scheduling following the optimized task execution order of 
workflow amid migrating workflow systems to the Kubernetes platform. 
This inconsistency in task scheduling order seriously degrades the efficiency of workflow execution and brings 
numerous challenges to the containerized process of workflow systems on Kubernetes.
In this paper, we propose a cloud-native workflow engine, also known as KubeAdaptor, a docking framework 
able to implement workflow containerization on Kubernetes, integrate workflow systems with Kubernetes, 
ensuring the consistency of task scheduling order. 
We introduce the design and architecture of the KubeAdaptor, elaborate the functionality implementation and 
the event-trigger mechanism within the KubeAdaptor.
Experimental results about four real-world workflows show that the KubeAdaptor ensures the consistency 
of the workflow systems and Kubernetes in the task scheduling order. 
Compared with the baseline Argo workflow engine, the KubeAdaptor achieves better performance in terms 
of the average execution time of task pod, average workflow lifecycle, and resource usage rate.
\end{abstract}

\begin{keywords}
 Workflow System\sep Containerization \sep Task Scheduling \sep Event Trigger \sep Kubernetes
\end{keywords}

\maketitle

\section{Introduction}
Cloud-native is considered the next future of cloud computing. 
Its presence has accelerated the technological revolution in the field of cloud 
computing~\cite{Gannon2017cloudnative}. 
As Kubernetes continues to win out in the container orchestration framework, 
the whole industry is embracing Kubernetes. 
The cloud-native technologies represented by containers, micro-services, DevOps (Development and Operations), 
and Kubernetes (K8s) reconstruct the IT operation, maintenance, and development mode, and also bring new 
opportunities for the rapid development of all industries in the cloud era~\cite{mao2020resource}. 
Presently, container technology and K8s have become mainstream tools for cloud resource 
management~\cite{bernstein2014containers} and dominated the whole cloud-native technology ecosystem.

Containers solve the problems of dependence and the operating environment's compatibility for software 
and provide a lightweight software packaging and distribution mechanism suitable for most cloud computing 
platforms~\cite{silver2017software}. 
As an excellent container orchestrator, the first hosted project by Cloud Native Computing 
Foundation (CNCF)~\cite{cncf2021}, K8s~\cite{k8s2021k8s} has become the de-facto standard container orchestration 
system. 
Nowadays, most production practice environments have migrated to the K8s platform with the prevalence of 
cloud-native technology. 
K8s and containers have become established standards for all cloud vendors, and the idea of cloud-based software 
development gradually takes shape. 
Traditional IT applications are speeding up the transition to cloud-native, and workflow system is no exception. 
Container technology and K8s provide a flexible mechanism for containerized execution of workflow 
systems~\cite{ren43826}.

Workflow systems often employ scheduling algorithms that optimize task execution order in a  
workflow to improve performance and execution efficiency~\cite{klop2018containerized}. 
Correspondingly, workflow scheduling algorithms implement workflow scheduling by adopting the optimal 
mapping method of tasks and resources. 
The cloud workflow instance obtains the mapping sequence of tasks and resources through scheduling algorithms, 
allocates the resources in the cloud to the workflow tasks, 
and finally completes the workflow execution process.
Workflow scheduling algorithms are core components of workflow systems. 
Recently, some researchers on workflow scheduling algorithms have focused
 on heuristic methods~\cite{kenari2021hyper}\cite{escott2020genetic}, 
 multi-objective optimization~\cite{chen2018multiobjective}, 
 deadline constraints~\cite{arabnejad2017scheduling}, 
 and other aspects~\cite{singh2018novel,chen2017real,psychas2019scheduling}, 
 aiming to find the best match of resources and tasks. 
However, the mapping method of tasks and resources in the workflow scheduling algorithms 
 aforementioned does not apply to the K8s production environment because 
 it ignores the inherent scheduler algorithm of the K8s. 
The integration of workflow systems and K8s follows a two-level scheduling scheme (\ref{sec:two-level}). 
It should be noted that the K8s scheduler has the characteristics of disordered scheduling, scattered scheduling, and 
unpredictability~\cite{menouer2021kcss}. These features resist workflow execution following the optimized task scheduling order 
obtained from the workflow scheduling algorithms, which fails to extend the efficacy of workflow scheduling 
algorithms to workflow containerized running on K8s and degrades the performance of the two-level 
scheduling scheme.
 
Many workflow systems are container-based and applied in specialized scientific areas, such as 
Pagasus~\cite{deelman2015pegasus}, Galaxy~\cite{jalili2020galaxy}, Taverna~\cite{wolstencroft2013taverna}, 
BioDepot~\cite{hung2019building}, Nextflow~\cite{nextflow2021}, Pachyderm~\cite{novella2019container}, 
Luigi~\cite{luigi2021}, SciPipe~\cite{lampa2019scipipe}, Kubeflow~\cite{kubeflow2021} and 
Argo workflows engine (Argo)~\cite{argo2021}. 
Most of them require the existence of the K8s cluster in cloud environments~\cite{ren43826}, which increases the 
research costs to some extent and limits the enthusiasm of researchers~\cite{rodriguez2020container}.
It would be more convenient for researchers to have a docking framework that could shield from the underlying K8s 
environment, go out of the box and allow users to run workflows in an engine-agnostic way.
So far, there is little work done in combining the workflow scheduling algorithms and K8s scheduler algorithm, 
and there is a lack of a docking specification and connection framework between both 
algorithms~\cite{klop2018containerized}. 
Furthermore, most works on workflow scheduling algorithms keep silent on workflow systems, stay away from the 
production environment, and only care about the simulation experiment environment CloudSim.
Therefore, we require a docking framework that bridges workflow systems or workflow scheduling algorithms 
to the K8s platform. It is meaningful and challenging work. 
It implements the effective fusion between workflow scheduling algorithms and K8s and enables users to 
carry out a flexible operation in K8s-based workflow systems.

In this paper, we propose KubeAdaptor for workflow systems on K8s. 
This docking framework addresses the inconsistent task scheduling order between workflow scheduling algorithms 
and K8s scheduler, enables workflows systems to be seamlessly migrated to K8s, and implements containerization of workflows. 
This framework redesigns the container creation functionality, 
utilizes the Informer component to monitor the underlying resource objects of K8s, 
and implements the logical structure of the whole framework using the Client-go package. 
In addition, this framework achieves data sharing between task containers 
through the dynamic volume mechanism of \verb|StorageClass|. 
This execution link completely integrates the workflow systems with the K8s scheduler.   
Experimental results about four real-world workflows show that our proposed KubeAdaptor achieves better 
performance in average task pod execution time, average workflow lifecycle, and resource usage rate.
Compared with the baseline Argo workflow engine, KubeAdaptor reduces average task pod execution time by up 
to $24.45\%$ (Montage), $47.57\%$ (Epigenomics), $23.72\%$ (CyberShake), and $24.65\%$( LIGO), average workflow 
lifecycle by up to $43.44\%$~(Montage), $43.65\%$~(Epigenomics), $44.86\%$~(CyberShake) and $48.98\%$~(LIGO), 
respectively. 
Our contributions are summarized as follows: 
\begin{itemize}
  \item Design a cloud-native workflow engine, which works as a docking framework to integrate workflow systems 
  with K8s. This docking framework implements workflow containerization on K8s platform while ensuring the consistency 
  of task scheduling under the two-level scheduling scheme (\ref{sec:two-level}).
  \item Implement a workflow injection module, the resource gathering module for experimental evaluation, 
  and an event trigger mechanism. 
  The workflow injection module is responsible for injecting workflow tasks into KubeAdaptor. 
  The resource gathering module is in charge of monitoring resource fluctuations and presents resource usage rates. 
  The event trigger mechanism optimizes the execution efficiency of each module within the KubeAdaptor through event invocation. 
  \item Provide a containerized solution with resource loads for workflow tasks to run four real-world 
  workflow applications and present the detailed performance analysis of KubeAdaptor compared to other workflow submission 
  methods.
  \end{itemize}

The rest of the paper is organized as follows. 
Section 2 introduces related work related to the topics in this introduction.
Section 3 presents the motivation, technological foundation, and implementation tools. 
Section 4 elaborates the KubeAdaptor design and shows its modular description, 
while section 5 further fulfills the experimental setup and evaluates the effectiveness 
of the KubeAdaptor. 
Finally, we summarize the paper in section 6.
We have open-sourced the KubeAdaptor. 
The source code is publicly available on GitHub at~\cite{kubeadaptor2021}.

\section{Related Work}
\label{sec:related}	
As a mainstream container orchestrator in the cloud-native era, K8s has won out over Mesos~\cite{hindman2011mesos} 
and Docker Swarm~\cite{dockerswarm2021} 
with ecology and technology advantages, elegantly decoupling application containers from the details of 
the system they run on. 
Before K8s as a mainstream container orchestrator, many researchers have attempted to harness container 
runtime for large-scale scientific computation.

Pegasus is a workflow management system for science automation, with scientific workflow portability and 
platform-agnostic descriptions of workflows. 
Recent work~\cite{vahi2019custom} incorporates a variety of container technologies in Pegasus to be suitable 
for varied execution environments. 
Galaxy is a data-analysis workflow platform that has gained more popularity by the development community 
in Bioinformatics. 
Through efforts from the development community, Galaxy supports offloading jobs in a variety of systems, 
ranging from Docker containers to other batch scheduler systems~\cite{gudukbay2021gyan}. 
Hung et al.~\cite{hung2019building} develop an open-source graphical workflow constructor, BioDepot, 
oriented to the field of bioinformatics engineering. 
It uses the Docker container to perform modular tasks while building workflow by dragging and dropping graphically. 
Nextflow is a data-driven computational pipeline for bioinformatics research developed by Barcelona Center 
for Genomic Regulation (CRG). 
It has been capable of using Docker and Singularity for multi-scale handling of containerized 
computation~\cite{di2017nextflow}. 
Airflow~\cite{airflow2021}, born in Airbnb and sponsored by Apache Incubator, is an open-source and distributed 
task scheduling framework based on Python programming. 
It is responsible for scheduling, monitoring, and managing workflows in a Docker container manner through 
DAG topology. 
Based on Makeflow, a tool for expressing and running scientific workflows across clustered environments, 
Albrecht et al.~\cite{albrecht2012makeflow} implement a batch system interface on Sun Grid Engine~(SGE) 
for scientific workflow. 
Afterward, Zheng et al.~\cite{zheng2015integrating} integrate Docker container runtime into Makeflow and 
Work Queue~\cite{bui2011work} for workflow scheduling.

These workflow systems using container technology significantly reduce the overhead of deploying custom 
computing environments and enable scientific workflow execution with scalability and reproducibility. 
However, intricate task dependencies within the workflow, data transmission and high concurrency among tasks, 
and coarse-granularity resource requirements, seriously degrade the performance of container-based workflow 
systems, resulting in task state detection delay, accessing bottleneck of shared data among tasks, 
garbage collection delay, inefficient resource detection, as well as poor fault-tolerant of 
containers~\cite{zheng2017deploying}. 
Instead, K8s elegantly addresses these issues mentioned above for different workflow systems by its 
inherent core concepts of \verb|Service|, \verb|Pod|, \verb|Volume|, \verb|Namespace|, 
and \verb|Deployment|. 
These concepts and methods make the K8s like a fish in water in terms of scheduling, automatic recovery, 
horizontal scalability, resource monitoring, and other aspects and go beyond the capabilities of 
container-based workflow systems.

On a different track, some workflow engines are continuously evolving with the prevalence of K8s. 
Through efforts within the PhenoMeNal H2020 Project~\cite{peters2019phenomenal}, Galaxy recently has gained 
the ability to deploy inside K8s via Helm Charts~\cite{moreno2019galaxy}.  
With the help of the K8s community, Nextflow provides built-in support for K8s, simplifying the execution 
of the K8s cluster containerized workflow~\cite{nextflow2021}. 
Nextflow may deploy the workflow execution as a K8s pod through the K8s executor. 
Presently, benefiting from the K8s ecosystem, the K8s Airflow Operator is released still under active 
development~\cite{airflow-operator2021}. 
The K8s Airflow Operator uses the K8s Python client to generate requests to the K8s apiserver and requires 
k8s to finish a series of operations related to workflow tasks in commercial clouds. 
Pachyderm~\cite{novella2019container} is a large-scale data processing tool natively built for running 
workflows on Docker and Kubernetes. 
It has also similarly enabled support for Kubernetes like Nextflow. 
Argo~\cite{argo2021}, an open-source project launched by Applatix and hosted by CNCF, provides a cloud-native 
workflow for K8s and implements each task in the workflow as a container. 
As the most popular workflow engine for K8s, Argo implements workflow functions for K8s through CRD 
(Custom Resource Definition). 
Kubeflow~\cite{kubeflow2021}, developed by Google, is another open-source platform for K8s, specifically 
dedicated to building and running machine learning workflows. 
Kubeflow provides a workflow tool called Kubeflow Pipelines based on the Argo workflow engine. 

These workflow engines always work with containers instead of command-line tools and use K8s as the orchestration 
framework in cloud environments. 
The integration between workflow systems and K8s improves the performance and efficiency of workflow execution, 
except for Pachyderm, Argo, and Kubeflow that are K8s-native and only support containers as means of processing. 
To our knowledge, apart from specialized techniques and built-in tools, the Galaxy, Nextflow, Airflow are 
all limited to the K8s scheduler and cannot determine the order in which task pods run. 
Additionally, these workflow systems require some service deployment before running workflows, 
which adds operational complexity. 
For K8s-native workflow systems like Argo, Pachyderm, and Kubeflow, the shortcoming is frequent access to 
the K8s cluster, posing excessive accessing pressure on the K8s cluster~\cite{chakraborty2020enabling}.
So we design and develop a cloud-native docking framework to efficiently implement workflow containerization. 
It only requires a few tweaks to the configuration file on deployment, greatly mitigates the pressure of accessing 
K8s apiserver with the help of Informer component, and enables users to run workflows on K8s without mental 
burden while ensuring the consistency of workflow scheduling algorithms and K8s scheduler.
In addition, the Galaxy, Nextflow, Airflow are not native workflow systems supporting K8s but have just 
transitioned to the K8s and are still working on features. 
Presently, the Argo is a cloud-native workflow engine designed for K8s, which has evolved into a generalized 
workflow engine. 
Furthermore, the core components of Kubeflow workflow processing are also based on Argo. 
Therefore, the experimental evaluation in this paper is only compared with Argo and the Batch Job submission 
approaches.




\section{Background}
In this section, we describe the motivation behind the proposed KubeAdaptor 
and introduce the technological foundation and tools that complement the implementation of this work very well.
\subsection{Motivation}
K8s, as you can see on its website, is an industrial-grade container orchestration platform. 
Its core functions are scheduling, auto-repair, horizontal scaling, service discovery, 
and load balance. 
\begin{figure}[htbp]
  \centering
  \includegraphics[width=\linewidth]{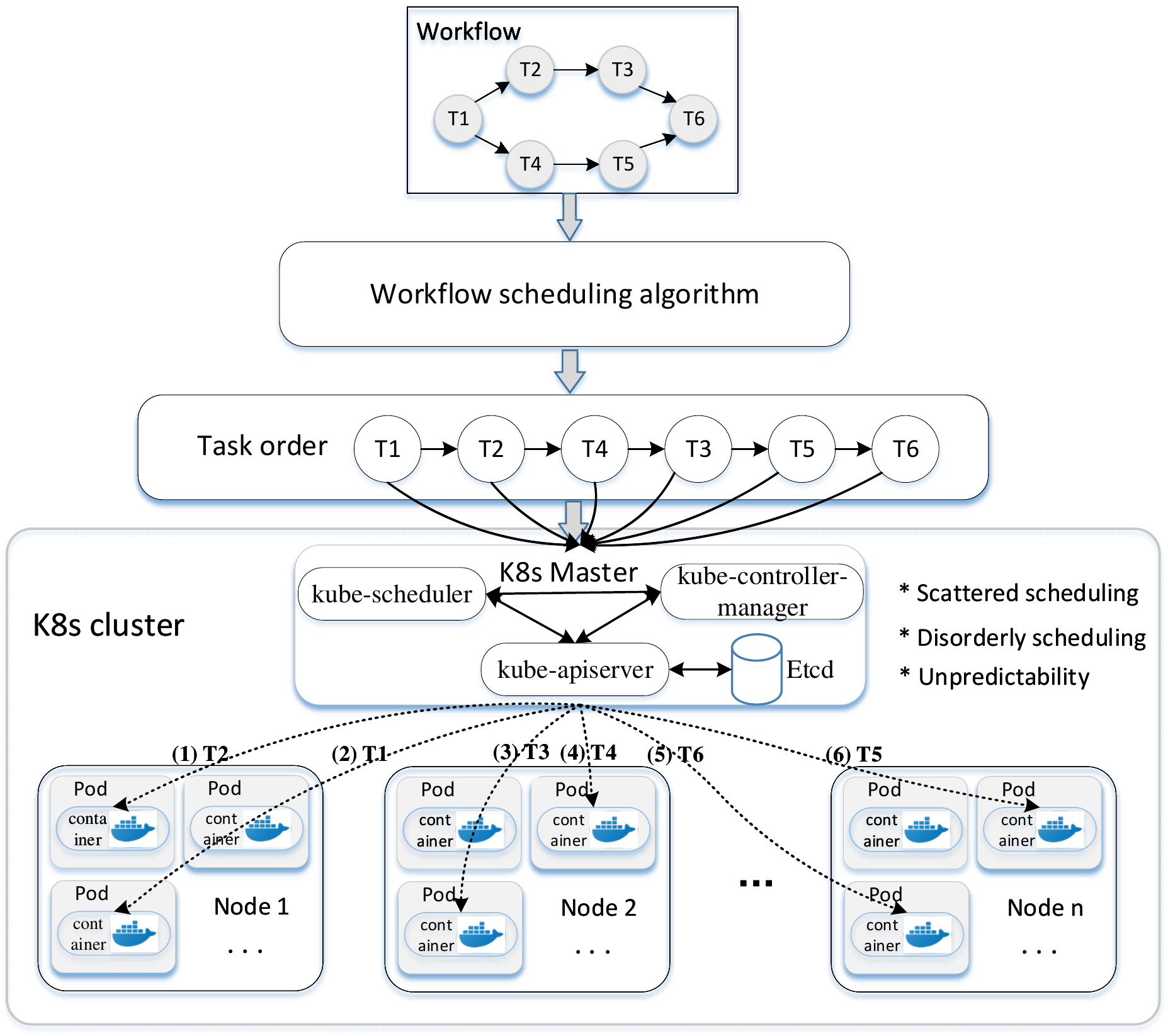}
  \caption{Task scheduling process graph in workflow containerization. 
  The serial numbers from (1) to (6) represent the execution order of task pods 
  scheduled by the K8s scheduler, which is inconsistent with the scheduling 
  results of the workflow scheduling algorithm and reflects the characteristics 
  of K8s scheduler scheduling strategy, such as scattered scheduling, 
  disorderly scheduling, and unpredictability.}
  \label{fig:scheduling}
\end{figure}
A \verb|Pod| is a minimum scheduling and resource unit of K8s. 
In workflow scheduling, tasks are encapsulated into pods and scheduled to run by K8s. 
K8s only cares about the scheduling and management of task pods 
rather than the context among task pods, which is the original intention of designing K8s. 
That is to say, the scheduling of task pods is doomed to be random scheduling and disorderly scheduling.
As shown in Figure.~\ref{fig:scheduling}, task order $\langle T1, T2, T4, T3, T5, T6 \rangle$  
is the scheduling results of the workflow scheduling algorithm under the two-level framework, 
followed by injection into the K8s cluster. 
The execution order of task pods scheduled by the K8s scheduler is 
$\langle T2, T1, T3, T4, T6, T5 \rangle$, 
which is inconsistent with this workflow scheduling algorithm, leading to the failure 
to obtain the superior performance of the workflow scheduling algorithm.

From this perspective, the K8s scheduler is unaware of the interdependencies among the tasks inside scheduled 
pods~\cite{klop2018containerized}. 
Due to the inconsistency between the task submission order and the K8s 
scheduling order, the K8s scheduler becomes an unpredictable and unreliable task scheduling method.
In addition, as mentioned in (\ref{sec:related}), the existing popular K8s-based workflow systems 
have complex deployment on various services and the excessive pressure of accessing K8s apiserver incurred by the 
frequent creation and destruction of \verb|Pod|, \verb|Namespace|, and \verb|PVC|, all of which weaken the performance 
of these workflow systems. 
To enable workflow systems to integrate the K8s platform smoothly and energize the two-level scheduling 
scheme, the KubeAdaptor is proposed to deal with these problems. 

\subsection{Technological Foundation and Tools}
In this subsection, we will introduce the technical foundation and tools. 
K8s resources related to the KubeAdaptor's implementation include \verb|Pod|, \verb|Service|, 
\verb|Namespace|, \verb|StorageClass|, etc.
\paragraph{\bfseries{Scientific Workflow.}}
\label{sec:scientificWf}
Workflow is an abstraction and automation of part or whole of a business process that defines tasks with 
interconnected dependencies and executes tasks by leveraging computing resources. 
A workflow, often described by a directed acyclic graph (DAG), 
represents a whole business process for user application. 
Accordingly, the dependencies between tasks are similar to the edges in 
a DAG diagram~\cite{zheng2017deploying}\cite{lee2013stretch}. 
Moreover, the task dependency is usually in the form of a shared file, 
created by one task and consumed by another.
\paragraph{\bfseries{Workflow Containerization.}}
Container technologies, such as Docker, a lightweight virtualization solution, 
encapsulate the workflow task running environment and required resources into a container~\cite{pahl2015containerization}. 
Moreover, the container possesses the characteristics of repeatability, reliability, and portability. 
These features allow users to focus on the task dependencies rather than the environment 
in which workflow tasks should run. 
In addition, containers have limited resource requirements, fast spin-up time, 
and low system overhead compared to virtual machines, so more containers can 
be accommodated in the same infrastructure as expected. 
Containers seem to be an ideal carrier for workflow tasks. 

These technological advantages of containers further foster workflow containerization. 
During the workflow containerization process, workflow tasks are packaged into containers 
through the Docker engine and built in the form of an {\itshape Image} file stored 
in local Harbor~\cite{harbor2021harbor} or remote Docker Hub repository~\cite{docker2021docker}. 
The task dependencies are essentially data dependencies that share data files 
among tasks through the storage volume of containers.

\paragraph{\bfseries{Informer.}}
Informer is the core toolkit in Client-go~\cite{clientgo2021}, 
responsible for synchronizing resource objects and events between K8s core components 
and Informer local cache. 
In brief, Informer uses a List-Watch mechanism to watch some resources, obtains 
the changes of these resources from the K8s apiserver, handles the changes of 
resources by the callback function registered by the user, and stores the change objects 
into the local cache persistently. 
In KubeAdaptor, we use the Informer toolkit to create monitoring objects \verb|podInformer|, 
\verb|nodeInformer|, and \verb|namespaceInformer| for the \verb|Pod|, \verb|Node|, 
and \verb|Namespace|, respectively. 
By watching the state changes of the task pod and workflow namespace, combined with 
the event trigger mechanism, this framework triggers the next task or workflow in real-time. 
The self-synchronized function between the local cache and the K8s apiserver 
ensures real-time updates of resource events. 
Through the Informer component, this framework can monitor the underlying resources 
of the K8s cluster, relieve the pressure of frequent access to the K8s apiserver,
 and improve its performance while ensuring the healthy operation on the K8s.

\paragraph{\bfseries{gRPC.}}
The gRPC is a high-performance open-source Remote Process Call (RPC) 
framework that can run in any environment. 
It wraps a service call in a local method, allowing the caller to invoke the service 
as if it were a local method, shielding it from implementation details. 
It can be applied in the data center, service scenarios within or across K8s clusters,
 and supports load balance, tracing, health checking, and authentication~\cite{gRPC2021}. 
The gRPC service requires a Protocol Buffer definition composed of a binary serialization 
toolset and language. 
This toolset can help the communication parties create a \verb|Proto| file with 
the communication data interface and generate the remote call functions 
required by the communication parties.
Using the gRPC framework, bi-directional streaming, and integrated authentication, 
we can quickly launch and scale to millions of RPCs per second. 
In this paper, We realize the information exchange between KubeAdaptor and 
the workflow injection module through gRPC communication. 
In this way, the workflows are injected into this framework from the workflow injection module.

\paragraph{\bfseries{Kubernetes Resource Objects.}}
With K8s apiserver as the core, the user interacts with the K8s system through 
the whole API set to perform \verb|CRUD| operations on cluster resources. 
End-users can often create, retrieve, update, and delete resources from the \verb|Kubectl| command 
line or directly from the RESTful API. 
The key K8s resources used in this work are as follows:
\begin{itemize}
\item \verb|Pod| is the atomic scheduling unit of K8s, a logical deployable unit, 
and a combination of multiple containers or many processes~\cite{verma2015large}. 
A pod is also a resource unit that defines how the container operates, 
such as command, environment variables, etc., and provides it with a 
shared operating environment, such as network and process space. 
For simplicity, the container where the workflow task resided is packed into a pod to participate in scheduling. 
\item \verb|Service| is an abstract way to expose an application running on a set of pods as a network service. 
It implements service discovery and load balance among K8s clusters. 
In K8s, the Service deploys the backend pods through Deployment, 
and this set of pods captured by a Service is usually determined by a selector 
in a deployable Yaml file. 
In this paper, the KubeAdaptor and the workflow injection module are deployed via Service. 
The backend behind Service runs the business pods of both modules, communicating via gRPC.
\item \verb|Namespace| is a way to divide cluster resources among multiple users. 
It provides scope for a namespace. Names of resources need to be unique within a namespace, 
but not across namespaces. 
Therefore, each workflow has its namespace with resource isolation in mind.
Tasks in the same workflow can access resources in this workflow namespace, 
not across namespaces, which ensures the operational security of the workflow. 
The workflow containerization process in KubeAdaptor firstly builds a namespace for the workflow, 
creates resources attached to this namespace, such as PersistentVolumeClaim (PVC), 
and then generates task pods. 
The activity domain of all task pods for this workflow is limited to this namespace.
\item \verb|StorageClass| works with PVC 
and external storage provisioner to provide dynamic storage for workflow namespaces. 
Each StorageClass contains the fields {\itshape provisioner}, {\itshape parameters}, 
and {\itshape reclaimPolicy}. 
These fields work when dynamically providing a PersistentVolume (PV) 
belonging to the StorageClass.
StorageClass needs to be deployed in the Master node via the Yaml file in advance. 
We provide dynamic NFS storage StorageClass template through the field {\itshape fuseim.pri/ifs} 
and deploy the NFS business mounting pod in a Deployment manner. 
When creating PVC in this workflow namespace code block, 
we set the field {\itshape storageClassName} to the name of StorageClass.
When creating workflow task pods in the code block of KubeAdaptor, 
we set the field {\itshape VolumeSource} to the PVC's name. 
Next, the NFS business mounting pod will mount the persistent volume 
for each task pod with the RBAC (Role-Based Access Control) permissions.
In this framework, multiple task pods within the same namespace share 
PV in a dynamic storage way through StorageClass and NFS provisioner.
\end{itemize}

\section{Design}
The main functionality of KubeAdaptor is to create the workflow namespaces and task pods, as well as 
the persistent storage volumes of pods, monitor the workflow namespaces and pod states through the Informer 
package, and trigger the following workflow or subsequent task in real-time in combination with the event 
trigger mechanism. 
In this section, we present a two-level scheduling scheme, elaborate workflow definition and the architecture 
of KubeAdaptor, and then introduce the workflow injection module, fault tolerance management, 
and event trigger mechanism.

\begin{figure}[h]
  \centering
  \includegraphics[width=3in]{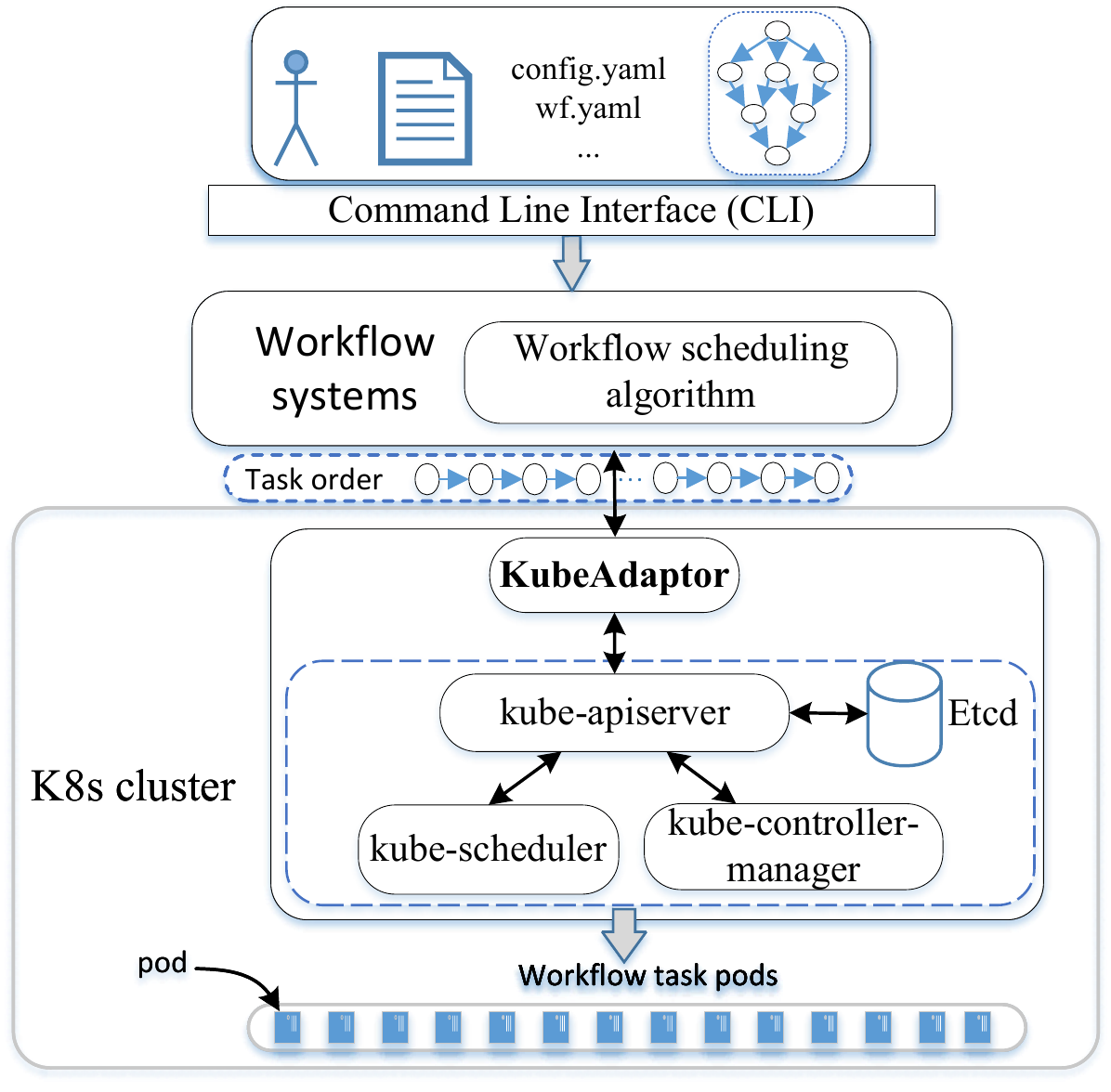}
  \caption{Two-level scheduling scheme. 
  }
  \label{fig:two-level}
\end{figure}
\subsection{Two-level Scheduling Scheme}
\label{sec:two-level}
The two-level scheduling scheme spans task scheduling of scheduling algorithms in workflow systems and 
pod scheduling of K8s. 
The integration of workflow systems and K8s follows the two-level scheduling scheme. 
As shown in Figure.~\ref{fig:two-level}, as soon as we deploy systematic configuration files and launch 
workflow systems through the command-line interface, the KubeAdaptor works as a logic interface and connects 
workflow scheduling algorithms to K8s. 
The task scheduling sequence optimized by the workflow scheduling algorithm is injected into KubeAdaptor 
by the workflow injection module (\ref{sec:inject}). 
The KubeAdaptor is responsible for finishing the containerization of injected workflow tasks on K8s. 
Meanwhile, the KubeAdaptor also ensures the consistency of task scheduling execution 
under the two-level scheduling scheme.

\subsection{Workflow Definition}
\label{sec:wf-define}
We use one-key deployment to run the KubeAdaptor and the workflow injection module to give users a good 
experience. 
Before deploying the workflow injection module and KubeAdaptor in containerized manner, we edit workflow 
definition and environmental deployment information one time via the Yaml file. 
ConfigMap decouples container images from variable configurations to ensure portability of workload pods. 
Variable configurations are defined in JSON format and describe the data definition of the workflow. 
ConfigMap uses the pod's volume to mount the configuration file to a specified directory in the container 
of the workflow injection module. 
This module deserializes the configuration file to obtain workflow information. 
Note that the ConfigMap and the workflow injection module container are defined in the same namespace.
\lstset{
     language = Pascal,
    basicstyle = \small\ttfamily,      
    breaklines = true,                  
    frame = single,                  
}
\begin{lstlisting}[caption =  The ConfigMap file for workflow injection module pod]
apiVersion: v1
kind: ConfigMap
metadata:
  labels:
    app: config
  name: dependency-inject
  namespace: default
data:
  dependency.json: |
    {
      "0": {
          "input": [],
          "output": ["1","2"],
          "image": ["shanchenggang/task-emulator:latest"],
          "cpuNum": ["1200"],
          "memNum": ["1200"],
          "args": ["-c","1","-m","100","-t","5"]
      },
      "1": {
          "input": ["0"],
          "output": ["3","4","5","6"],
          "image": ["shanchenggang/task-emulator:latest"],
          "cpuNum": ["1200"],
          "memNum": ["1200"],
          "args": ["-c","1","-m","100","-t","5"]
      }, 
      . . .
\end{lstlisting}
Listing 1 shows the definition of workflow information in ConfigMap by the Key-Value method. 
The {\itshape Key} and the {\itshape Value} refer to the name and content of the variable configuration file, respectively. 
The variable configuration file is defined in JSON format and contains the definition of workflow task nodes. 
Each task node works as a step of a workflow and contains \verb|input| attribute, \verb|output| attribute, 
\verb|image| attribute, \verb|cpuNum| attribute, \verb|memNum| attribute, and \verb|args| attribute.
Six attributes of the task node represent input dependencies, output dependencies, 
task image address, CPU requirement, memory requirement, and running parameters for this task 
pod, respectively. 
The steps in a workflow are executed following the order defined as DAG. 
As shown in Listing 1, workflow definition in KubeAdaptor is human-readable and extremely simple 
to use with a negligible learning burden.
\begin{figure}[h]
  \centering
  \includegraphics[width=\linewidth]{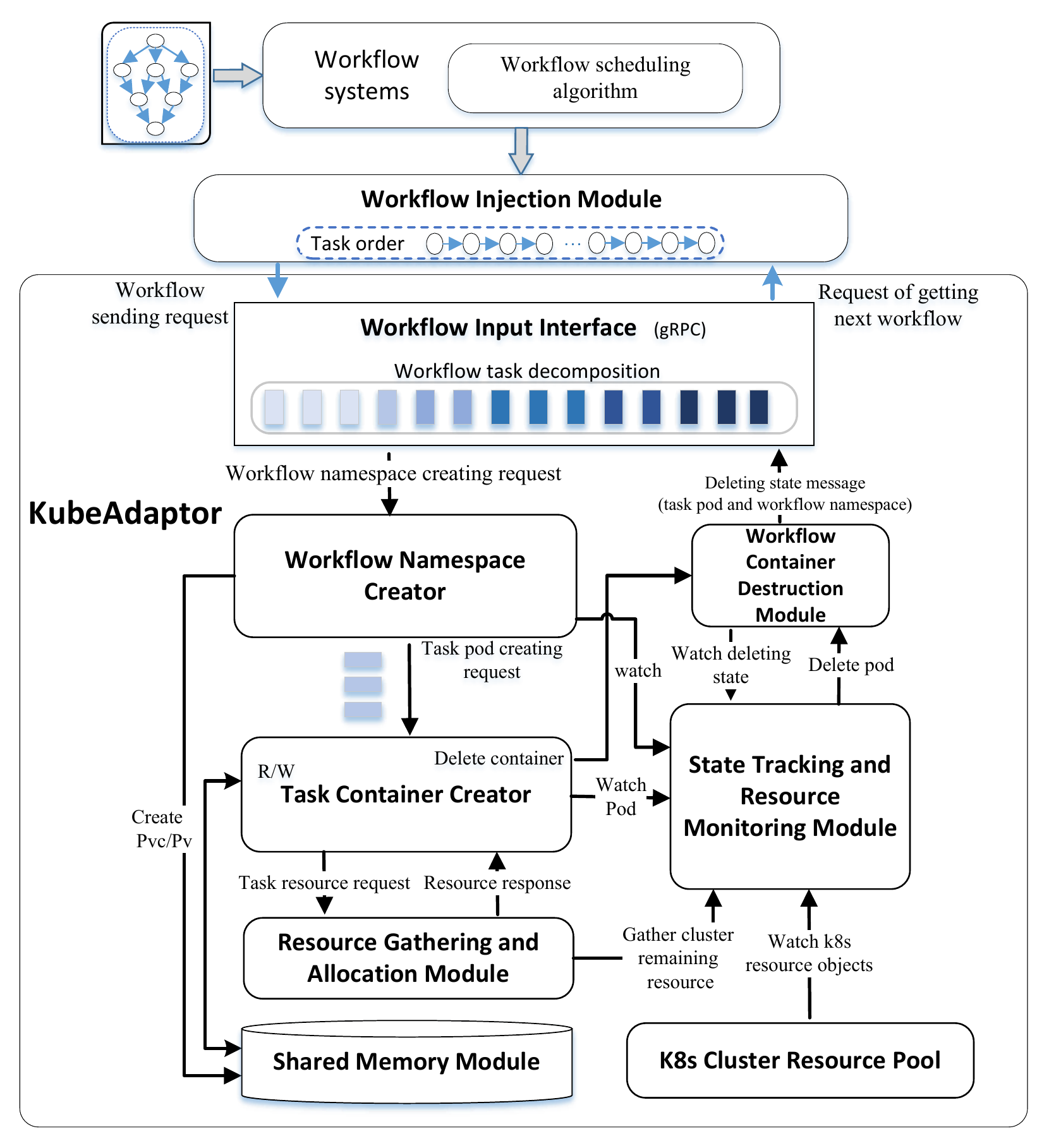}
  \caption{Architecture diagram of the KubeAdaptor. 
  This diagram illustrates the internals of KubeAdaptor in detail. 
  The workflow injection module is independent of KubeAdaptor and responsible for injecting the workflow 
  task sequence into the KubeAdaptor. 
  }
  \label{fig:cwb-arch}
\end{figure}

\subsection{The Architecture of KubeAdaptor}
\label{sec:cwb-architecture}
The KubeAdaptor consists of several components that talk to each other during workflow containerization. 
The key components of KubeAdaptor are elaborated in detail throughout this section. 
The architecture of KubeAdaptor is shown in Figure.~\ref{fig:cwb-arch}.

\paragraph{\bfseries{Workflow Input Interface.}}
As soon as the KubeAdaptor and the workflow injection module are deployed and launched, 
this module starts to receive the workflow generation request from the workflow injection module via 
{\itshape gRPC}, decomposes the workflow tasks, and turns to the workflow namespace creator module.
 Meanwhile, this module watches the state changes of the task pods or workflows just created 
 and triggers the subsequent execution steps at any time.

If the state of the current task pod is \verb|Failed|, the system is abnormal, and this module turns to 
fault tolerance management~(\ref{sec:fault}) to repair. 
If the state of the current task pod is \verb|Succeeded|, this module invokes an event triggering scheme 
and triggers the workflow container destruction module to remove the completed task pod.  
Next, when detecting a successful return flag of the removed task pod, the event trigger scheme 
immediately sends the generation request of the subsequent task pod to the workflow namespace creator 
module. 
In addition, if the current workflow is completed, as soon as this module captures a successful 
removal flag of the completed workflow, 
it uses the event callback function to send a request of acquiring the subsequent workflow to the 
workflow injection module.

\paragraph{\bfseries{Workflow Namespace Creator.}}
This module creates the workflow namespaces against requests of generating workflows to realize resource 
isolation between different workflows. 
This module first obtains a list of \verb|NamespaceLister| resources by state tracking and resource monitoring module 
and checks whether or not the namespace to be created exists. If this namespace does not exist, this module creates 
this namespace, uses the \verb|StorageClass| to dynamically generate the PVC that meets the capacity 
requirements of PV for different workflows. 
Instead, This module proceeds to the task container creator module.

\paragraph{\bfseries{Task Container Creator.}}
This module is responsible for generating task pods under namespace of the specified workflow. 
This module first obtains the list of \verb|PodLister| resources under the namespace by 
invoking the state tracking and resource monitoring module and checks 
whether or not the requested pod under the namespace exists.
If the requested task pod exists, this module turns to fault tolerance management~(\ref{sec:fault}). 
Instead, this module allocates the resource to create the task pod and sets the volume source of this task pod 
to PVC's name in this namespace. 
Next, the task pod can operate on the PV to read and write data 
when mounting NFS shared directory successfully. 
This module uses the Goroutine mechanism to create concurrent task pods against parallel offspring tasks.  

\paragraph{\bfseries{Resource Gathering and Allocation Module.}}
\label{sec:resource-gathering}
This module is in charge of requesting the resource allocation of the building task pods. 
First, this module obtains the resource list of \verb|NodeLister| and \verb|PodLister| in the K8s cluster 
by invoking the state tracking and resource monitoring module. 
Second, this module acquires the resource amount occupied by all pods through the field {\itshape Requests}
of the pods in \verb|PodLister| and gets the total allocatable resource amount 
of the cluster by the field {\itshape Allocatable} of all \verb|NodeLister| nodes 
except for the Master node. 
The difference between the above two is the number of remaining resources available for allocation in the K8s 
cluster. 
Finally, this module allocates the requiring resources to the task container generation 
module with the overall resource requirements of the global workflows and 
resource allocation algorithm in mind.

\paragraph{\bfseries{State Tracking and Resource Monitoring Module.}}
This module is essentially a monitor program based on the List-Watch mechanism running in the backend. 
It provides \verb|PodLister|, \verb|NodeLister|, and \verb|NamespaceLister| resource lists for 
other modules, and responds to resource monitoring requests of each module at any time. 
This module mainly monitors the execution states of the workflow task pods and workflows, and feeds back to 
the workflow input interface module. 

\paragraph{\bfseries{Workflow Container Destruction Module.}}
This module responds to the workflow input interface module and performs the pod deleting operation invoked 
by event trigger mechanism. 
In addition to deleting the \verb|Succeeded| or \verb|Failed| pods, the namespace of the completed 
workflow is also included.
This module uses the API deletion method of the Client-go package~\cite{clientgo2021} to implement 
the deletion operation. 

\paragraph{\bfseries{Shared Memory Module.}}
The shared memory approach is a solution to implement data dependence between workflow tasks. 
This module works with NFS sharing between cluster nodes and requires deploying the NFS business 
mounting pod and StorageClass service in advance. 
First, the KubeAdaptor creates PVC according to user requirements combined with StorageClass service. 
Then the KubeAdaptor generates PV through an external NFS business mounting pod and the PVC. 
The created task pod binds PV and PVC and implements data sharing via a mounted sharing directory.
Multiple task pods within the same namespace use this shared directory to exchange data, 
which implements the data dependence of task nodes described in the DAG diagram.

\subsection{Workflow Injection Module}
\label{sec:inject}
The workflow injection module is an auxiliary module independent of KubeAdaptor, which facilitates the 
implementation of our framework. 
It is responsible for reading variable configuration information of workflow definition, 
parsing and generating workflows, responding to input requests of the subsequent workflow, 
and injecting workflow information into KubeAdaptor via gRPC. 
The following components are key parts of the workflow injection module.
\paragraph{\bfseries{Workflow Parser.}}
This module takes care of reading variable configuration files from the mounted directory inside the pod of 
the workflow injection module, deserializes the JSON file composed of workflow definition, 
and encapsulates workflow information. 

\paragraph{\bfseries{Workflow Sending Module.}}
When the KubeAdaptor starts to work, this module finishes the initialization of the workflow, 
obtains the Service IP address and port number of KubeAdaptor's pod 
through the environment variable, remotely accesses the KubeAdaptor, and sends the workflow. 
All the subsequent sending operations of this module are performed under the trigger of KubeAdaptor's pod. 

\paragraph{\bfseries{Next Workflow Trigger Module.}}
This module enables launching the gRPC server locally, responds to the request from KubeAdaptor's pod, 
and invokes the workflow sending module to send the subsequent workflow.

\subsection{Fault Tolerance Management}
\label{sec:fault}
This module deals with the exceptions of the workflow input interface modules and the task container 
creation module. 
The failure of task pod creation is mainly due to the unsuccessful mounting of the NFS service for 
the task pod. 
Once the task pod creation fails, this module will enable the event trigger mechanism to call back 
the task pod creation function. 
It sends the failed task pod generation request to the workflow namespace creator module again till 
this task pod is successfully created.  
Generally, as long as the K8s cluster remains healthy and stable and the NFS business pod is up and 
running, the KubeAdaptor has no multiple task failure cases.
In addition, if the task pod to be created already exists in its namespace, 
this module will enable the task container creation module to throw an exception and 
call back the workflow container destruction module to remove it via an event trigger scheme. 
Then this module invokes the task container creator module to request the generation of the exception 
pod again.

\subsection{Event Trigger Mechanism}
The event triggering mechanism embodies the interaction process of KubeAdaptor internal components. 
It uses the Informer component and the Event package to implement the event callbacks and respond 
to the state changes of various resources in K8s. 
When launching the event triggering mechanism, the KubeAdaptor utilizes the event package to register 
key events and responds to state changes of resources in real-time through the callback function. 
For example, when a task pod is done, the KubeAdaptor fetches the state changes of this task pod 
through the Informer component in real-time, invokes the deleting operation of this task pod and 
triggers the generation of the subsequent task pod. 

This mechanism enables a quick switch between the creation and destruction of workflow task pods 
and restricts the out-of-order pod scheduling of the K8s scheduler.
Through the event triggering mechanism and the KubeAdaptor's core function, 
the K8s scheduler can schedule the workflow task pods as the workflow scheduling algorithms expected 
while being aware of the state changes of various resources in K8s. 
This mechanism ties the components within KubeAdaptor together and speeds up the execution efficiency 
of its internal modules. 
\begin{figure*}[h]
  \centering
  \includegraphics[width=\linewidth]{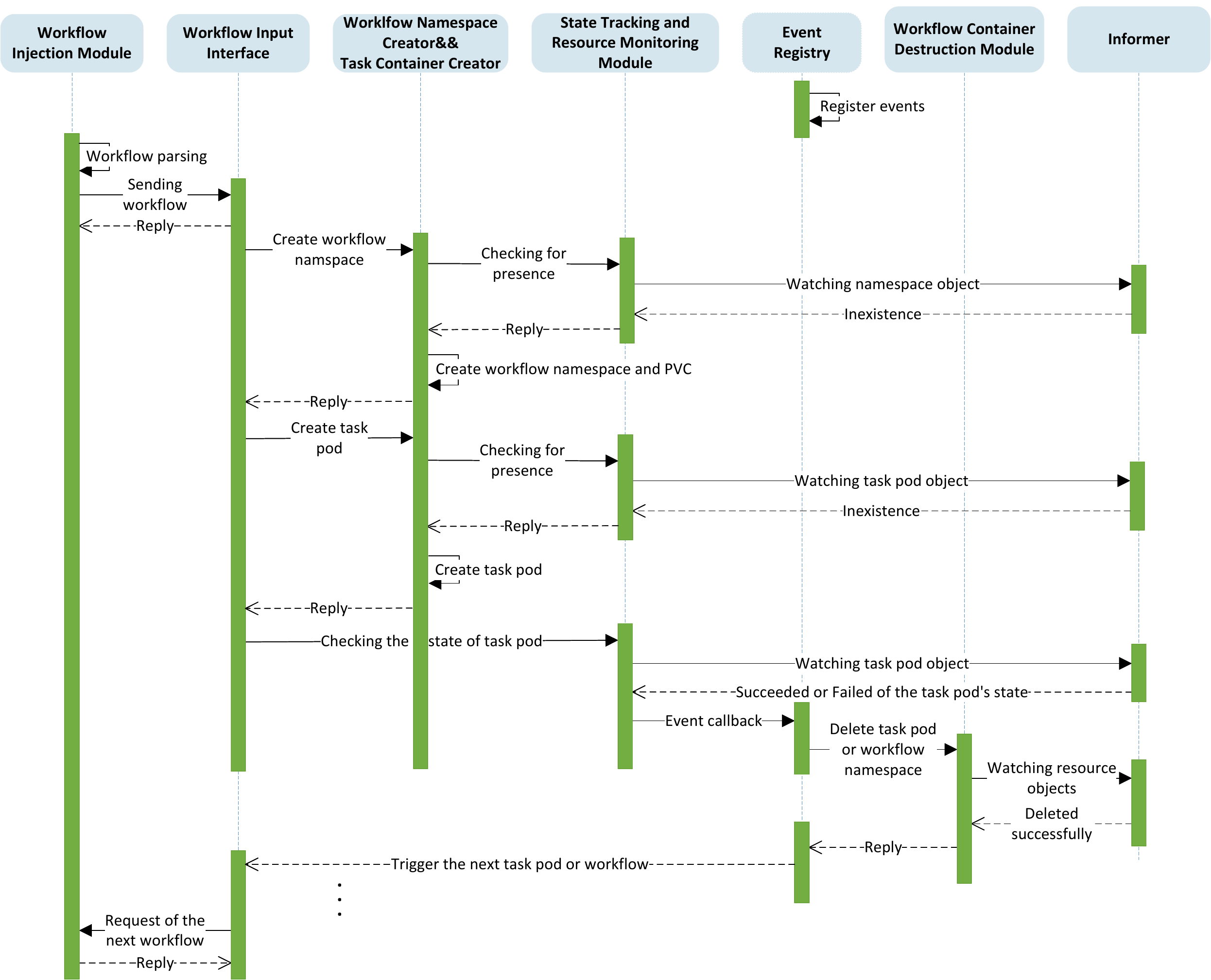}
  \caption{Sequence diagram of the event trigger mechanism. 
  The execution sequence of the workflow task, obtained from the workflow scheduling algorithm, 
  is injected into the workflow input interface of KubeAdaptor. 
  This figure depicts the interactions between the components within the KubeAdaptor to 
  implement the event triggering mechanism. 
  }  
  \label{fig:event}
\end{figure*}
As shown in Figure. \ref{fig:event}, the event registry is in charge of registering key events 
by Event package and finishing event callback. 
The registered event callback behavior function runs in the backend. 
First, the workflow injection module injects workflow information into the KubeAdaptor. 
Then, the workflow input interface receives the requests of sending workflow and outputs 
requests of creating task pods and workflow namespace. 
Next, the workflow namespace creator module generates workflow namespace, and the task container creator 
creates task pods. 
Once the state tracking and resource monitoring module watches task pods with 
\verb|Failed| or \verb|Succeeded| state, this module enables the event 
callback function to invoke workflow container destruction module to do relevant deletion operations. 
It should be noted that the fault tolerance management function is also incorporated.
In addition, the state tracking and resource monitoring module monitors resource state changes 
by interacting with the Informer component. 
Accordingly, under the action of event triggering mechanism, the workflow container destruction 
module coordinates the state tracking and resource monitoring module with the workflow input interface 
to respond to the corresponding operations described in the KubeAdaptor architecture 
(\ref{sec:cwb-architecture}).

\section{Experimental Evaluation}	
\subsection{Experimental Setup}
\label{sec:experimental}
To evaluate the efficacy of the KubeAdaptor, we design the workflow injection module. 
This module and the KubeAdaptor are containerized and deployed into the k8s 
cluster through {\itshape Service} and {\itshape Deployment}. Both modules 
communicate with each other through the gRPC mechanism.
The Image extraction policy of the workflow task is set to the {\itshape PullifNotPresent} field.  
The KubeAdaptor starts the state tracking and resource monitoring module at runtime, 
which monitors the \verb|Namespace|, \verb|Pod|, and \verb|Node| 
resources of the K8s cluster in real-time and invokes the event trigger module at any time. 
We explore the performance of three workflow submission approaches by running four real-world 
workflow applications on the K8s.
The K8s cluster consists of one master node and six nodes. Each node possesses 
an 8-core AMD EPYC 7742 2.2GHz CPU and 16GB of RAM, 
running Ubuntu 20.4 and K8s v1.19.6 and Docker version 18.09.6.

\subsection{Workflow Example}
\begin{figure*}
  \centering
  \includegraphics[width=6.8in]{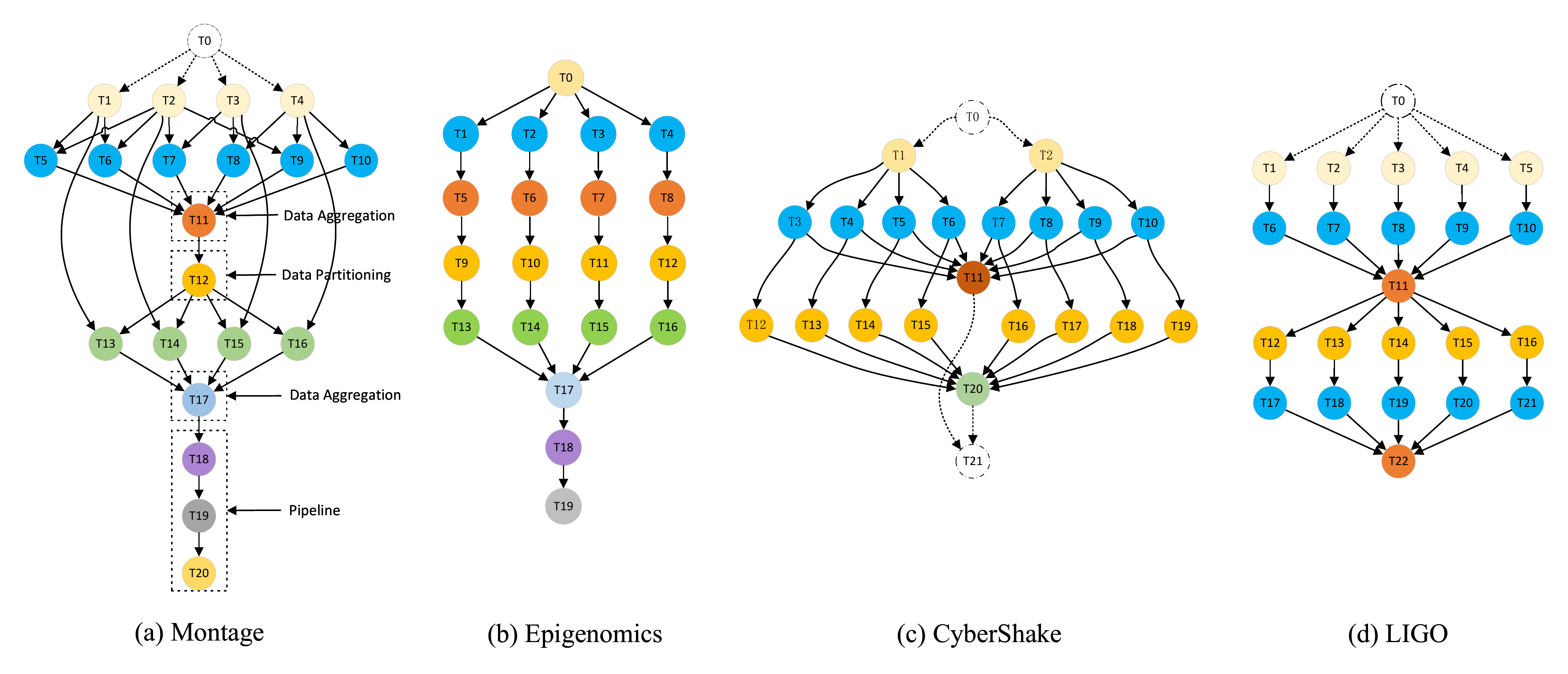}
  \caption{The topology diagram of four real-world workflow applications.}
  \label{fig:four-topology}
\end{figure*}

To verify the application scalability of the KubeAdaptor, we employ four classes of scientific workflow 
applications, such as Montage~(astronomy), Epigenomics~(genome sequence), CyberShake~(earthquake science), 
LIGO Inspiral~(gravitational physics)~\cite{juve2013characterizing}. 
We add virtual task nodes at the entrance and exit of workflows to facilitate the construction of the DAG 
workflow structure. 
These four types of workflows cover all the elementary structural characteristics concerning composition 
and components (in-tree, out-Tree, fork-join, and Pipeline) that can validate the scalability of KubeAdaptor. 
For each class of workflow, the workflow structure with a smaller task size (about 20) is selected in 
the experiment, as shown in Figure~\ref{fig:four-topology}, extracted from the Pegasus Workflow 
repository~\cite{pegasus2021}.
We assume that four classes of scientific workflows have the same task and the workflow task program 
uses resource loads to simulate workflow tasks in the experiments. 
In each workflow topology, a node represents a workflow task. 
A directed edge between nodes represents a task dependency. 
According to the relationship among task nodes, the scheduling algorithm of these workflow schedules tasks 
topologically in a top-down fashion.

\begin{figure*}
\centering
\subfigure[Montage]{\label{fig:subfig:a}
\includegraphics[width=0.45\linewidth]{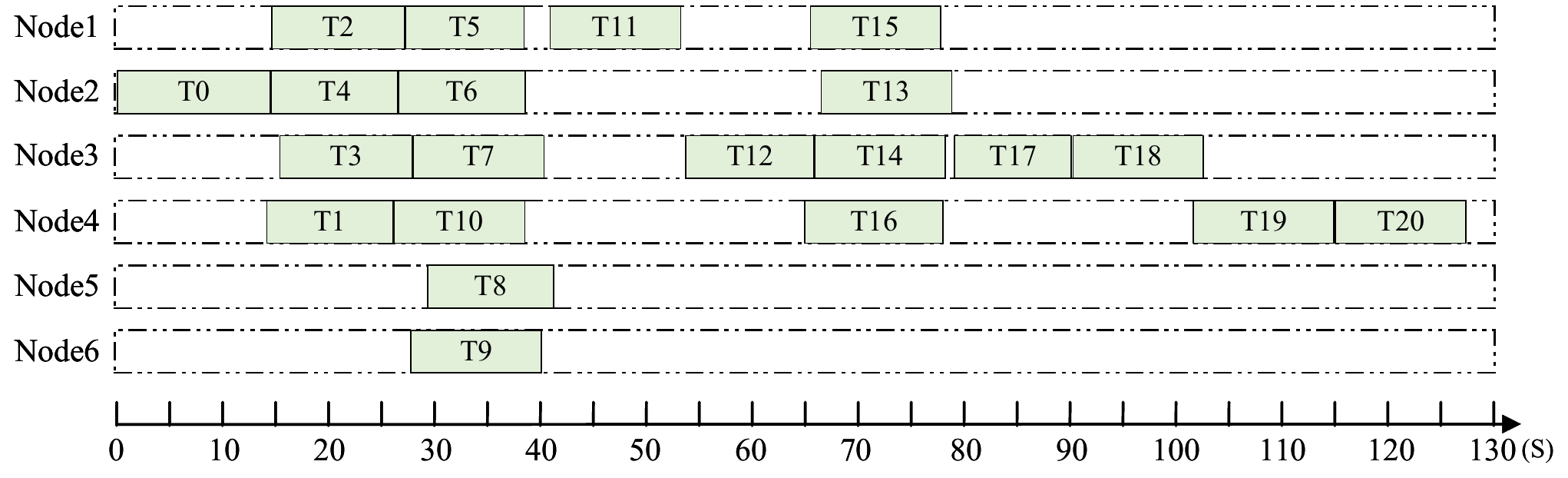}}
\hspace{0.01\linewidth}
\subfigure[Epigenomics]{\label{fig:subfig:b}
\includegraphics[width=0.45\linewidth]{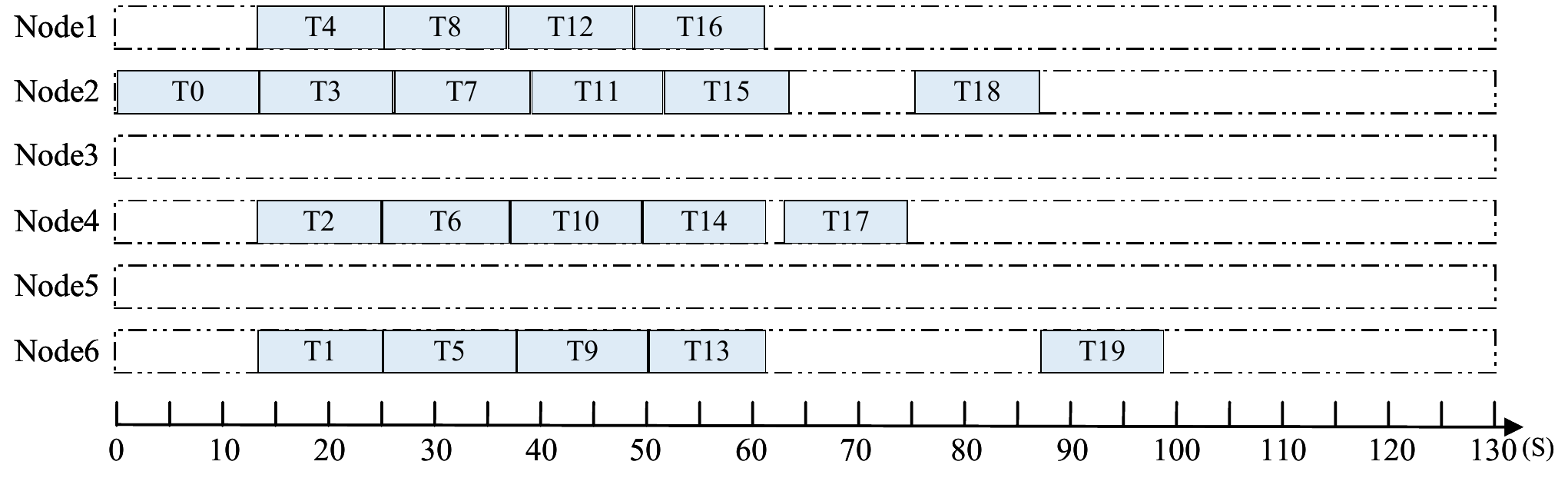}}
\subfigure[CyberShake]{\label{fig:subfig:c}
\includegraphics[width=0.45\linewidth]{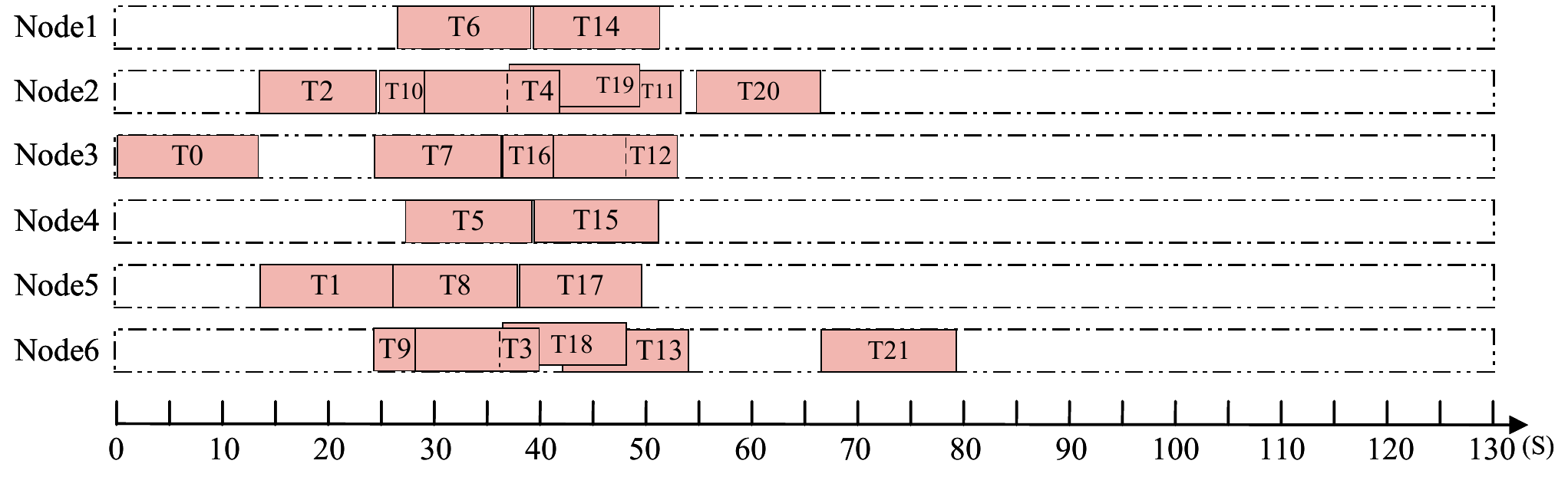}}
\hspace{0.01\linewidth}
\subfigure[LIGO]{\label{fig:subfig:d}
\includegraphics[width=0.45\linewidth]{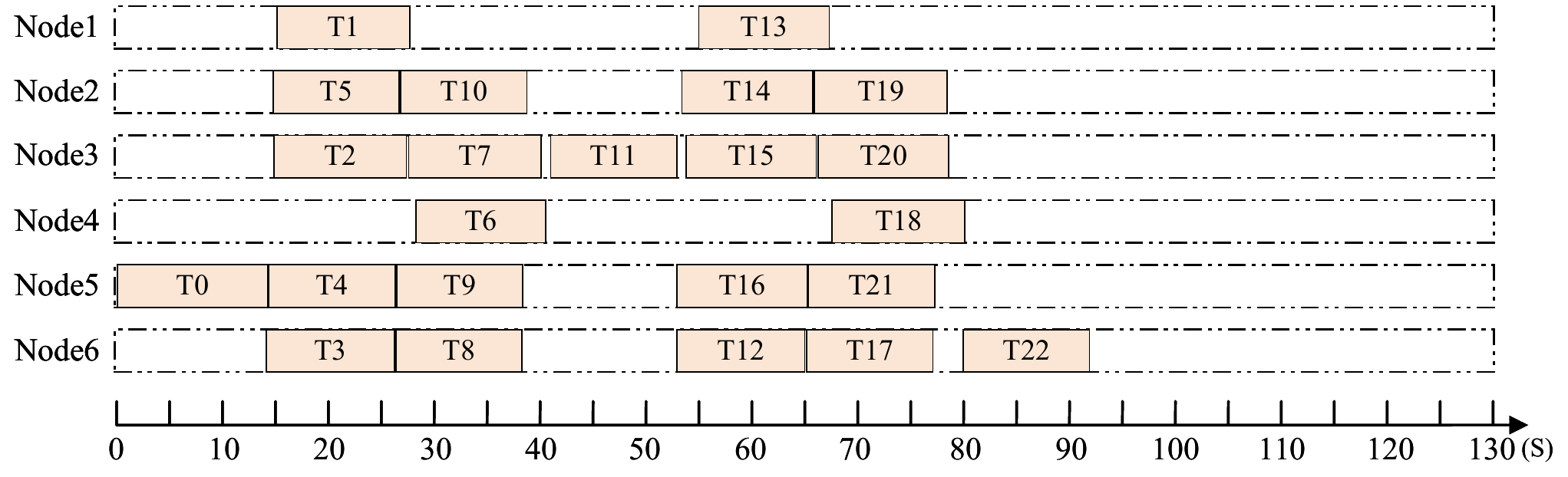}}
\caption{Schedules of four real-world workflows in Figure~\ref{fig:four-topology} with (a) Montage (workflow lifecycle = 127.129s), 
(b) Epigenomics (workflow lifecycle = 99.182s), (c) CyberShake (workflow lifecycle = 78.939s), and (d) LIGO (workflow lifecycle = 92.361s).}
\label{fig:sequence}
\end{figure*}

\paragraph{\bfseries{Workflow Instantiation.}}
We design a {\itshape Python} application as the main program of a workflow task. 
This {\itshape Python} application uses the {\itshape Stress} 
tool~\cite{stress2021} to emulate CPU and memory usage in a given time~\cite{klop2018containerized}. 
We pack the Python application into a task Image file by Docker engine~\cite{taskemulator2021}, 
store the task Image file in local Harbor~\cite{harbor2021harbor} or remote Docker Hub repository~\cite{docker2021docker}, 
and initialize the task Image address in the ConfigMap file of the workflow injection module. 
We use the ConfigMap method to inject container parameters into the task container. 
Task container parameters include CPU cycles, memory allocation, and the duration 
of the task pod~(\ref{sec:wf-define}).

In the main program of the task pod, we use several resource parameters to instruct {\itshape Stress} to work. 
Each task uses the {\itshape Stress} tool to set $1$ CPU fork, 
a memory of $100MB$, and a duration of $5$ seconds. 
CPU forking and memory allocation operations in the task pod last $10$ seconds in total. 
In the deployment Yaml file of KubeAdaptor, we set the resource {\itshape requests} and resource 
{\itshape limits} parameters for the task pod to $1200$ Milli cores (i.e.,$1200m$) CPU and $1200Mi$ memory. 
Note that the {\itshape requests} field has the same parameters as the {\itshape limits} field.

\subsection{Results and Analysis}
In this subsection, we first verify the consistency of task scheduling order 
between the workflow scheduling algorithms and the K8s scheduler through KubeAdaptor. 
Then we compare the proposed KubeAdaptor with Batch Job submission, Argo, in terms of workflow 
execution efficiency, CPU usage rate, and memory usage rate. 
Since Argo is a general cloud-native workflow engine in the industry, 
we use it as a baseline workflow engine in our experiments. 
The following is a description of the three workflow submission methods.
\begin{itemize}
  \item KubeAdaptor: We employ the containerized method to deploy the KubeAdaptor, as mentioned 
  in (\ref{sec:experimental}).
  \item Batch Job: We use a customized {\itshape Shell} script to submit workflow tasks with Job type in 
  batches via the {\itshape Kubectl} command. 
  \item Argo: We define the workflow task dependency relationship as DAG, described via {\itshape Yaml} file, 
  and submit the {\itshape Yaml} file to Argo workflow engine through the Argo binary tool~\cite{argo2021}. 
\end{itemize}

\paragraph{\bfseries{Scheduling Order Consistency Analysis.}}
We set the Master node not to participate in pod scheduling so that the Master node can bear less resource load 
and keep K8s as healthy as possible, which is in line with the strategy of excluding the Master node in 
the resource gathering module (\ref{sec:resource-gathering}).
To verify the performance of KubeAdaptor, we set the image pull policy to the {\itshape ifNotPresent} 
field in the deployment Yaml file and pull task images from local Harbor. 
After we deploy {\itshape Yaml} files in the K8s cluster, 
KubeAdaptor and workflow injection module are scattered and scheduled to the cluster nodes in the form of a pod, 
and both of them communicate via gRPC. 
Once the deployment is successful, the workflow injection module sends workflows to the KubeAdaptor via gRPC. 
When receiving workflows, the KubeAdaptor starts the workflow containerization process.

As shown in Figure.~\ref{fig:sequence}, we run four real-world workflows to obtain its scheduling sequences. 
Task scheduling orders in each subfigure are strictly consistent with its respective scheduling sequence of 
topology diagrams in a top-down fashion.
Each workflow lifecycle includes creating workflow namespace, PVC, task pod, and deleting task pod and 
workflow namespace throughout scheduling timelines.
When the number of concurrent tasks in the workflow is greater than the node number in the K8s cluster, 
some nodes in the K8s cluster will host multiple task pods simultaneously within resource capacity. 
Instead, some nodes in K8s cluster will be idle, as shown in Figure.~\ref{fig:subfig:c} and Figure.~\ref{fig:subfig:b}, 
respectively.

Each task program uses the {\itshape Stress} tool to last $10s$ through forking CPU 
and allocating memory. 
We use a set of workflow lifecycle sample data from the experiment to depict the 
entire workflow lifecycle timeline. 
The workflow lifecycles of four real-world workflows in Figure.~\ref{fig:sequence} from creation to death are 
$127.129s$~(Montage), $99.182s$~(Epigenomics), $78.939s$~(CyberShake), and $92.361s$~(LIGO), respectively. 
We can observe that each workflow task is completed separately throughout the respective workflow lifecycle. 
In addition, the concurrent execution of multiple tasks significantly improves the workflow execution efficiency 
of the KubeAdaptor.
It verifies that KubeAdaptor is instrumental in integrating workflow systems with the K8s. 
\paragraph{\bfseries{Workflow Lifecycle Comparison.}}
\label{sec:performance}
We adopt four real-world workflows shown in Figure.~\ref{fig:four-topology} and run it 100 times through 
KubeAdaptor, Batch Job, and Argo, respectively. 
We can obtain the average execution time of each task pod and the whole workflow lifecycle in each real-world 
workflow on K8s for three workflow submission approaches.

\begin{figure}[h]
  \centering
  \includegraphics[width=\linewidth]{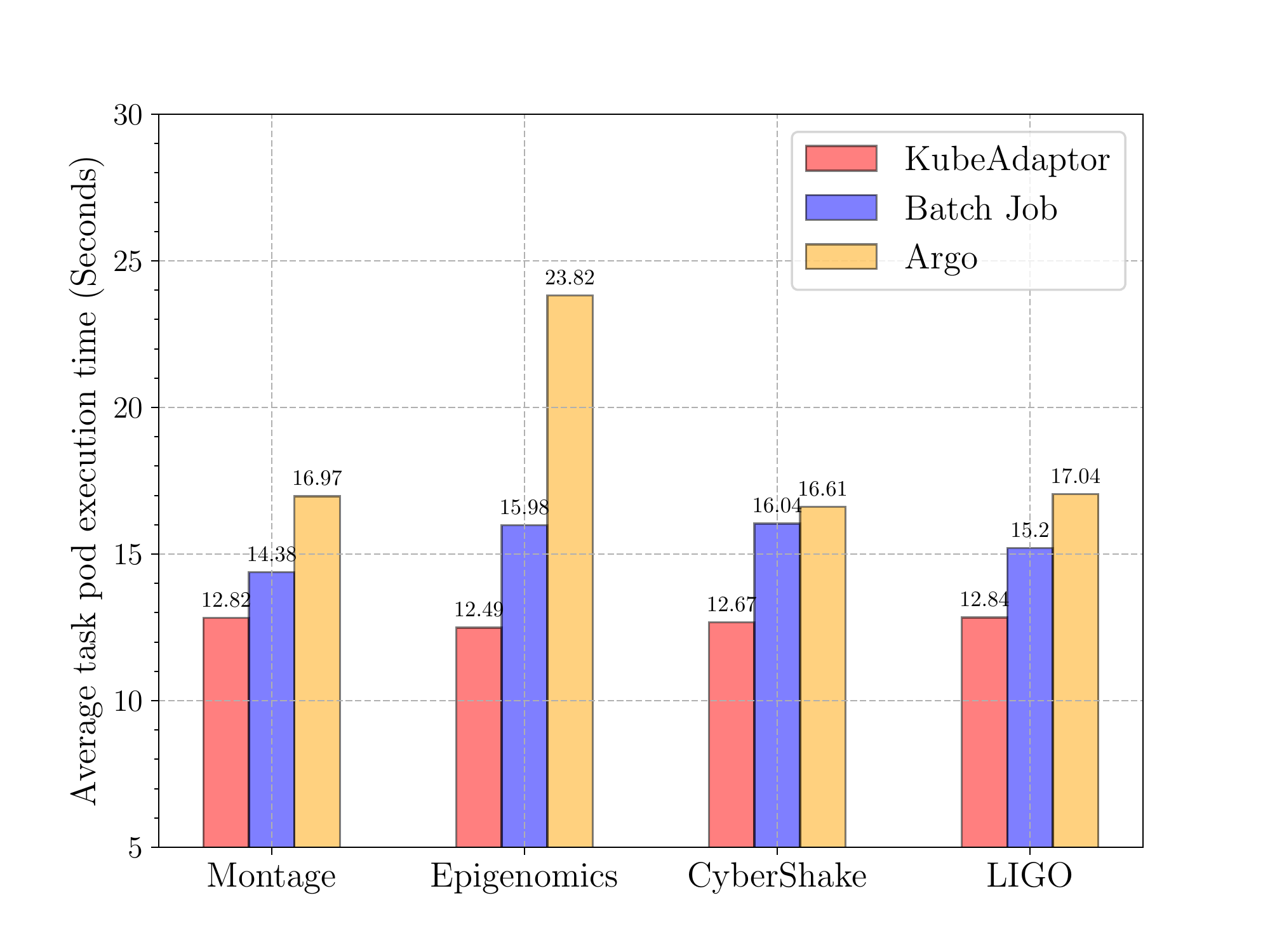}
  \caption{Average execution time of task pod. 
  The execution time of the task pod is the elapsed time from creation to deletion.}
  \label{fig:aveargeTaskTime}
\end{figure}
Figure.~\ref{fig:aveargeTaskTime} shows the average execution time of task pod in four real-world workflows 
for three distinct workflow submission approaches. 
We select a workflow lifecycle with an average workflow execution time from $100$ experiments and obtain 
the average execution time of the task pod in this workflow. 
As shown in Figure.~\ref{fig:aveargeTaskTime}, in four real-world workflows, the KubeAdaptor obtains $12.82s$, $12.49s$, $12.67s$, 
and $12.84s$ in terms of the average execution time of the task pod, respectively, 
and significantly outperforms Batch Job and Argo.
In terms of the average execution time of the task pod, the KubeAdaptor reduces by $24.45\%$, $47.57\%$, 
$23.72\%$, and $24.65\%$ in four real-world workflows, respectively, 
compared with the baseline Argo workflow engine.

\begin{figure*}[h]
    \centering
    \includegraphics[width=\linewidth]{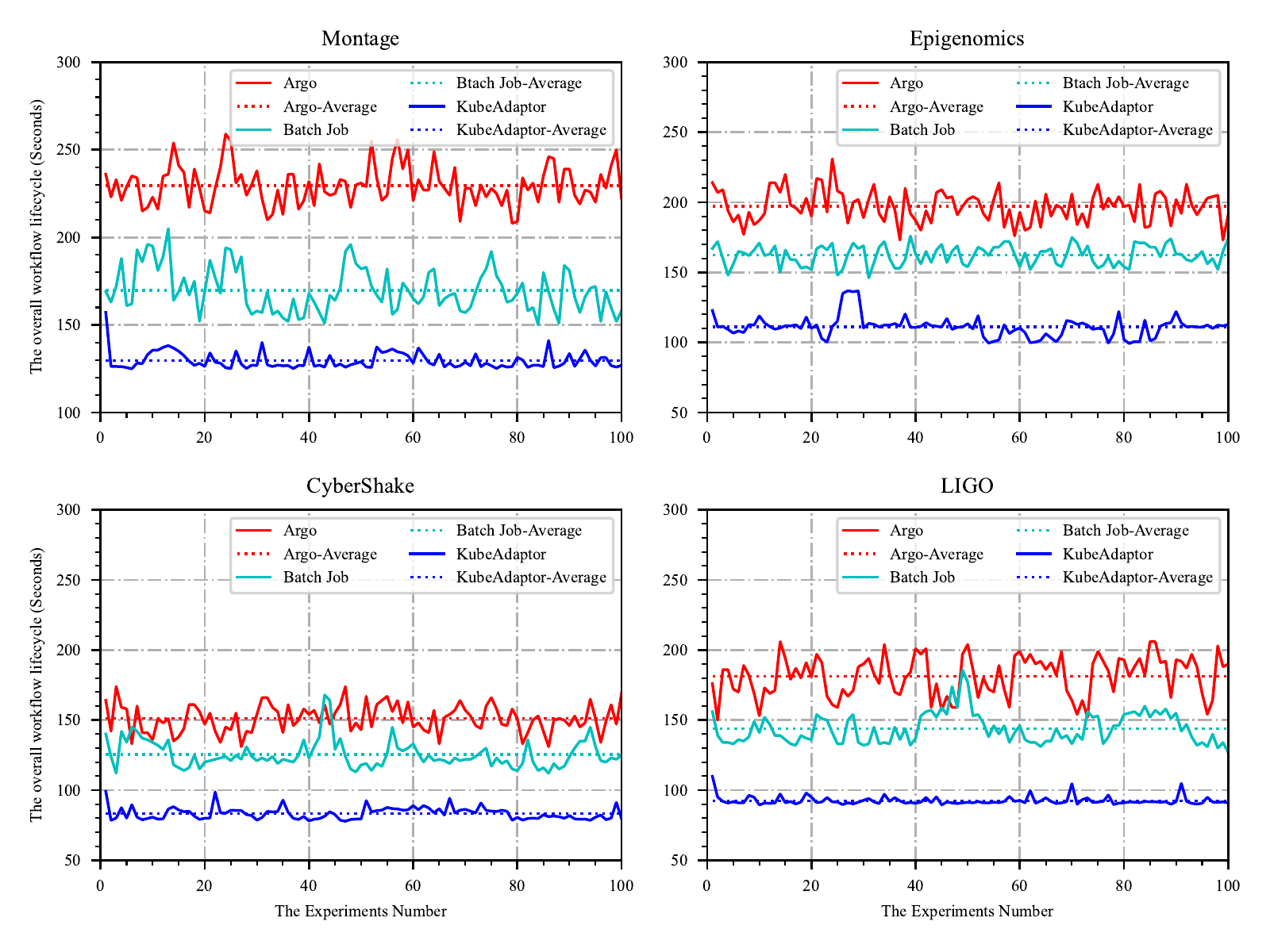}
    \caption{Average workflow lifecycle. 
    Workflow lifecycle refers to the entire process from creation to death of workflow namespace. 
    On the aspect of average workflow lifecycle, this figure contains four subgraphs, such as Montage~($129.85s$ of KubeAdaptor, 
    $169.83s$ of Batch Job, and $229.57s$ of Argo), Epigenomics~($111.12s$ of KubeAdaptor, $162.34s$ of Batch Job, 
    and $197.18s$ of Argo), CyberShake~($83.36s$ of KubeAdaptor, $125.44s$ of Batch Job, and $151.19s$ of Argo), 
    and LIGO~($92.46s$ of KubeAdaptor, $143.8s$ of Batch Job, and $181.22s$ of Argo).
    }
    \label{fig:averageWfTime}
  \end{figure*}
Further analysis shows that Argo does not perform well in average task pod execution time due to its greater 
focus on the depth of the workflow topology and the number of workflow tasks. 
For the Batch Job approach, the {\itshape Kubectl} command deployment pattern spends too much time 
during the interacting process with the K8s apiserver. 
Compared to the Argo and Batch Job, for each task pod in workflow, 
the KubeAdaptor saves some time by reconstructing container generating functionality. 

\begin{figure*}
\centering
\subfigure[Montage]{\label{fig:cpu25000:a}
\includegraphics[width=0.45\linewidth]{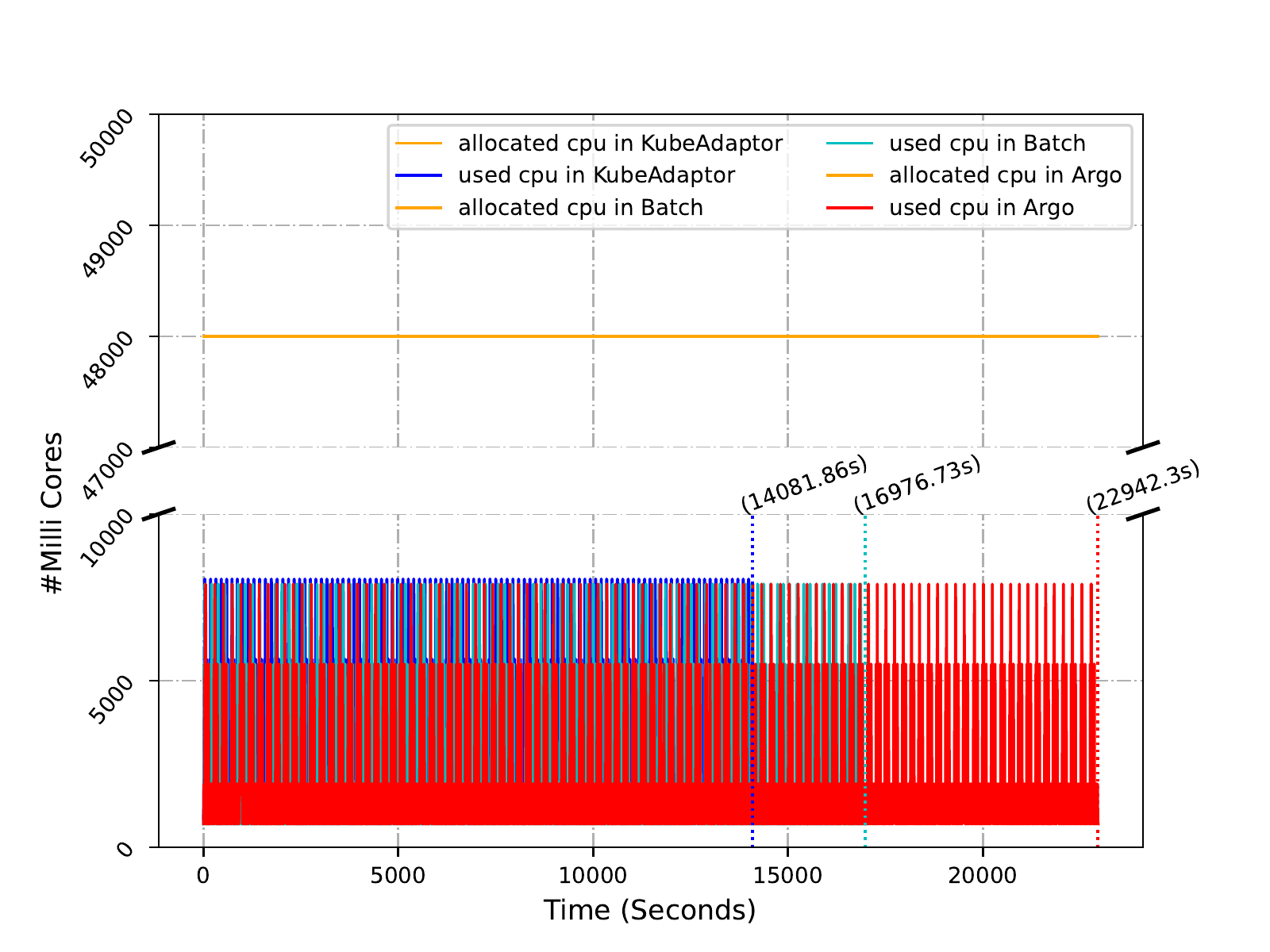}}
\hspace{0.01\linewidth}
\subfigure[Epigenomics]{\label{fig:cpu25000:b}
\includegraphics[width=0.45\linewidth]{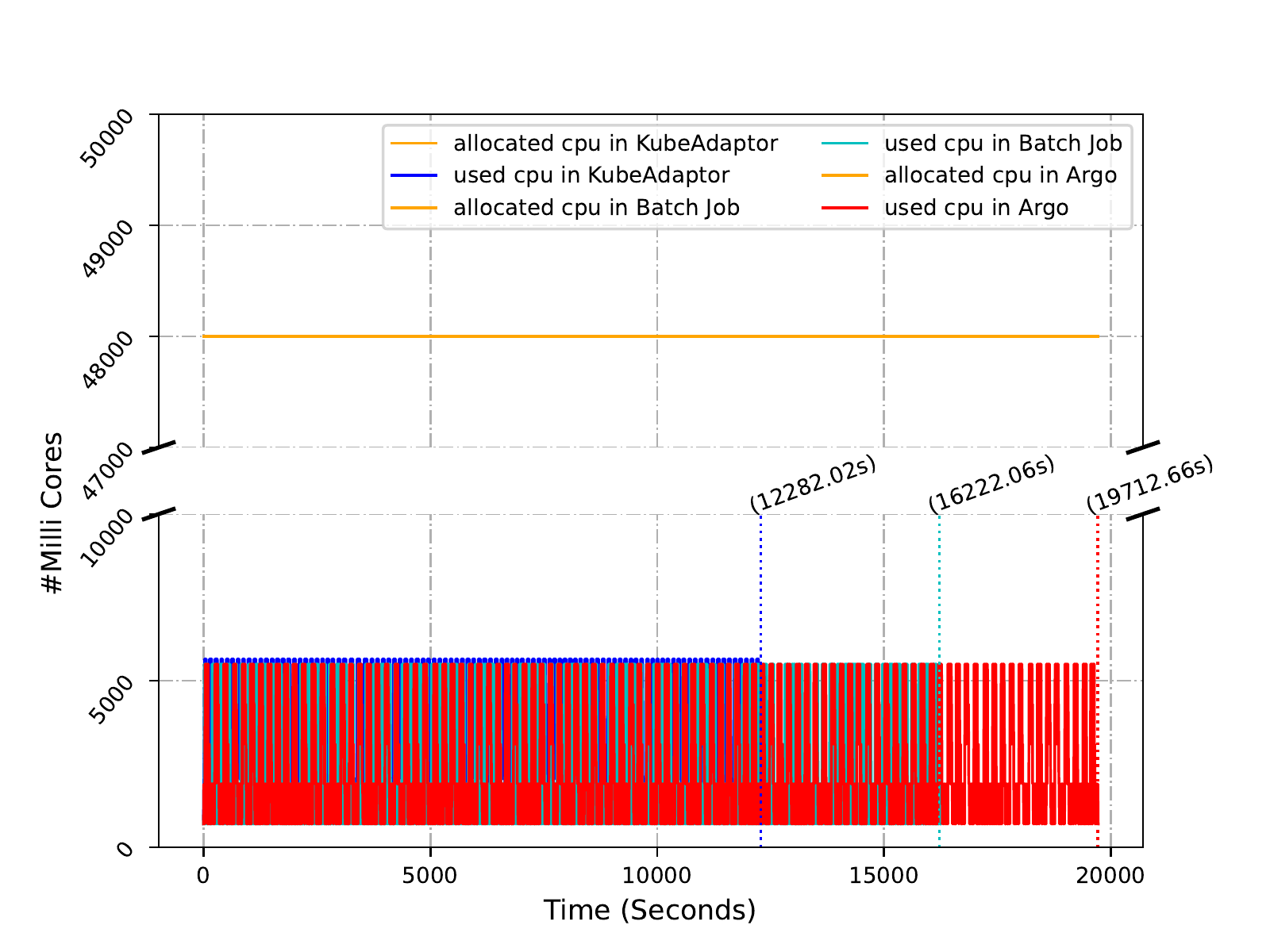}}
\subfigure[CyberShake]{\label{fig:cpu25000:c}
\includegraphics[width=0.45\linewidth]{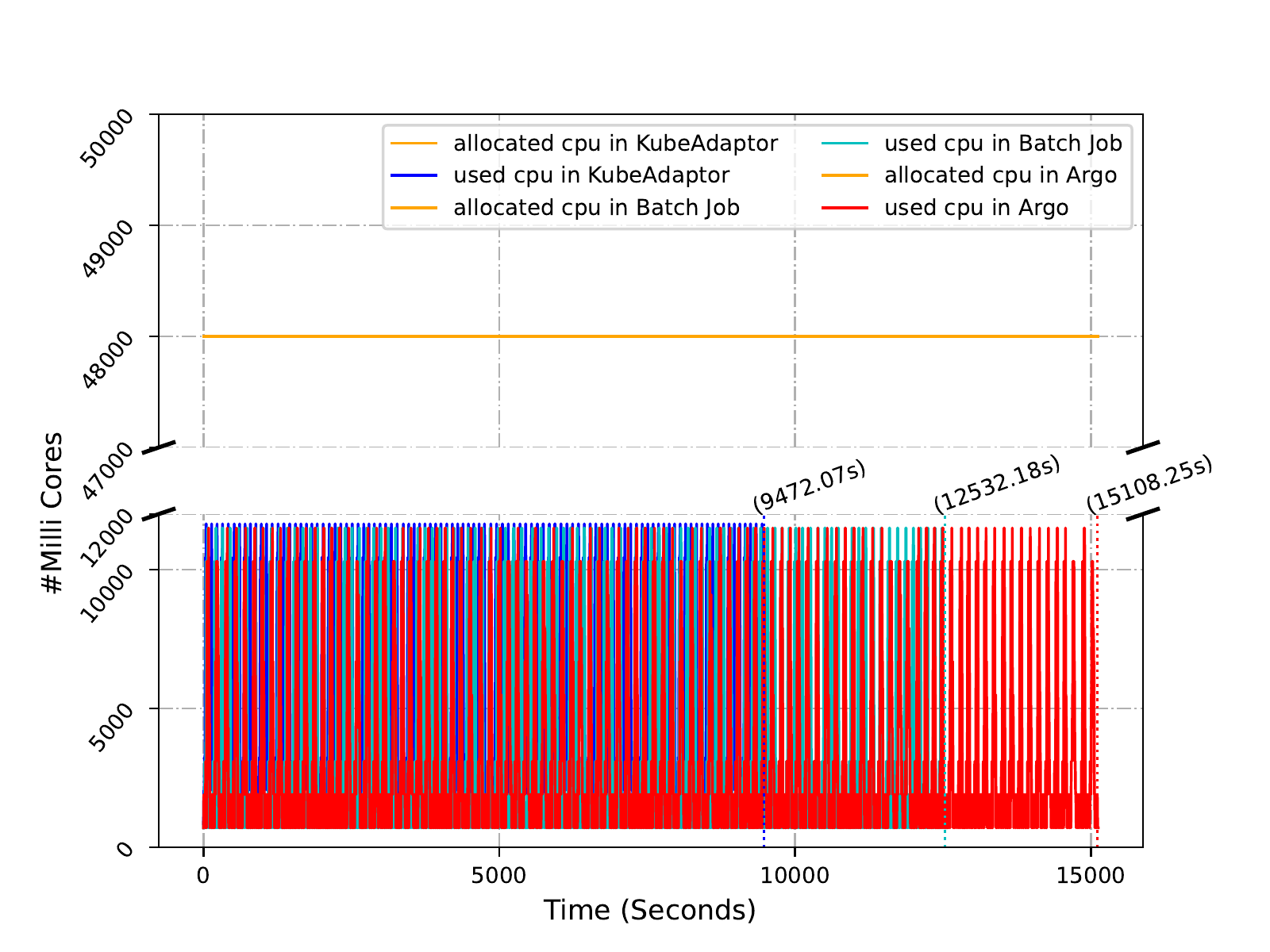}}
\hspace{0.01\linewidth}
\subfigure[LIGO]{\label{fig:cpu25000:d}
\includegraphics[width=0.45\linewidth]{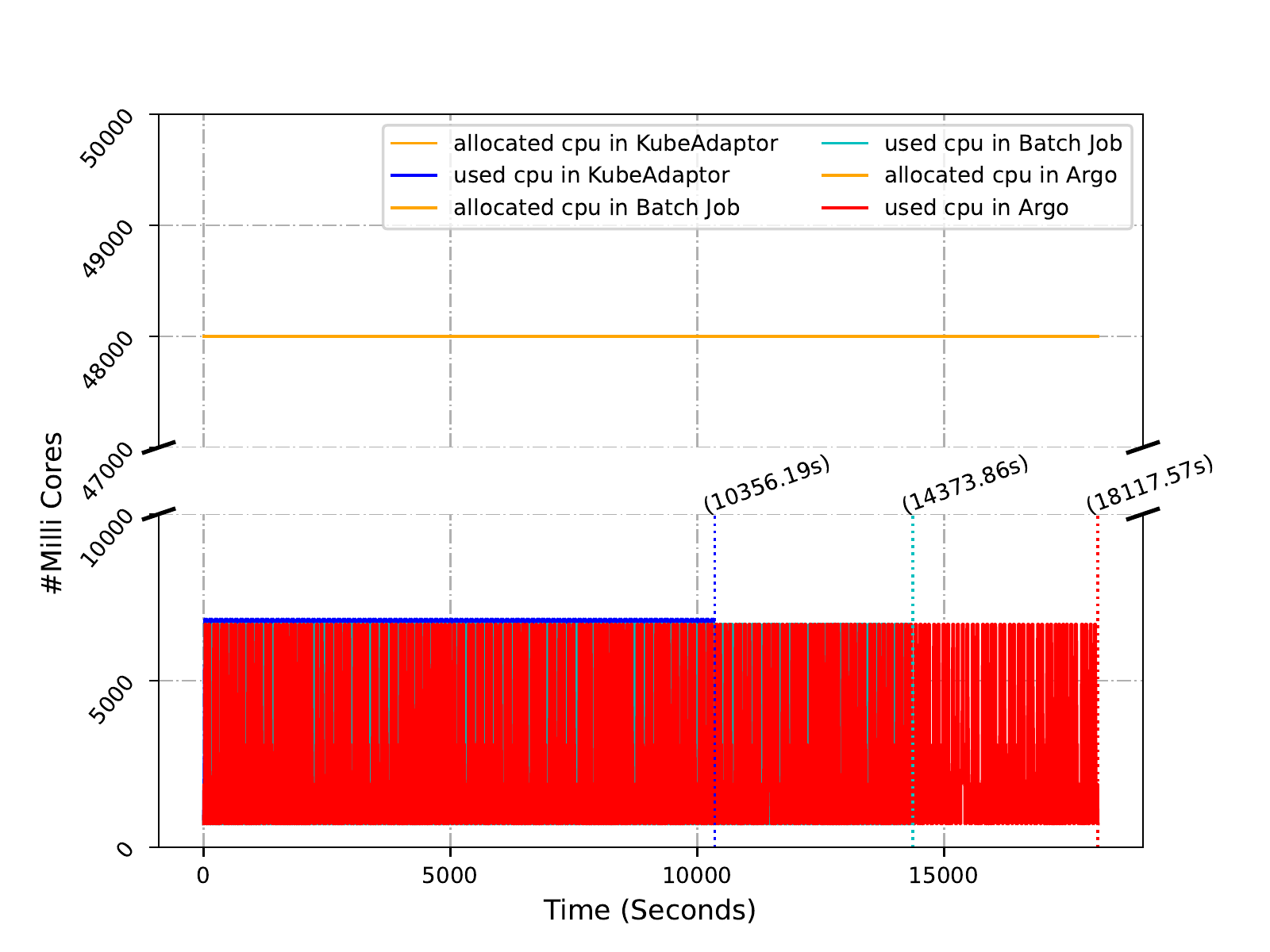}}
\caption{
The CPU usage rate of four real-world workflows in $100$ consecutive experiments. 
The four subfigures show the curve of CPU resource usage of four real-world workflows under 
three workflow submission approaches, respectively.  
Except for the Master node, the other nodes participate in resource load and have $48000m$ of CPU in all. 
The orange curve in each subfigure depicts the number of allocatable CPU in the K8s cluster. 
The blue, cyan, and red curves depict the number of CPU used in KubeAdaptor, Batch Job, and Argo, 
respectively.
}
\label{fig:cpu25000-usage}
\end{figure*}

\begin{figure*}
\centering
\subfigure[Montage]{\label{fig:mem25000:a}
\includegraphics[width=0.45\linewidth]{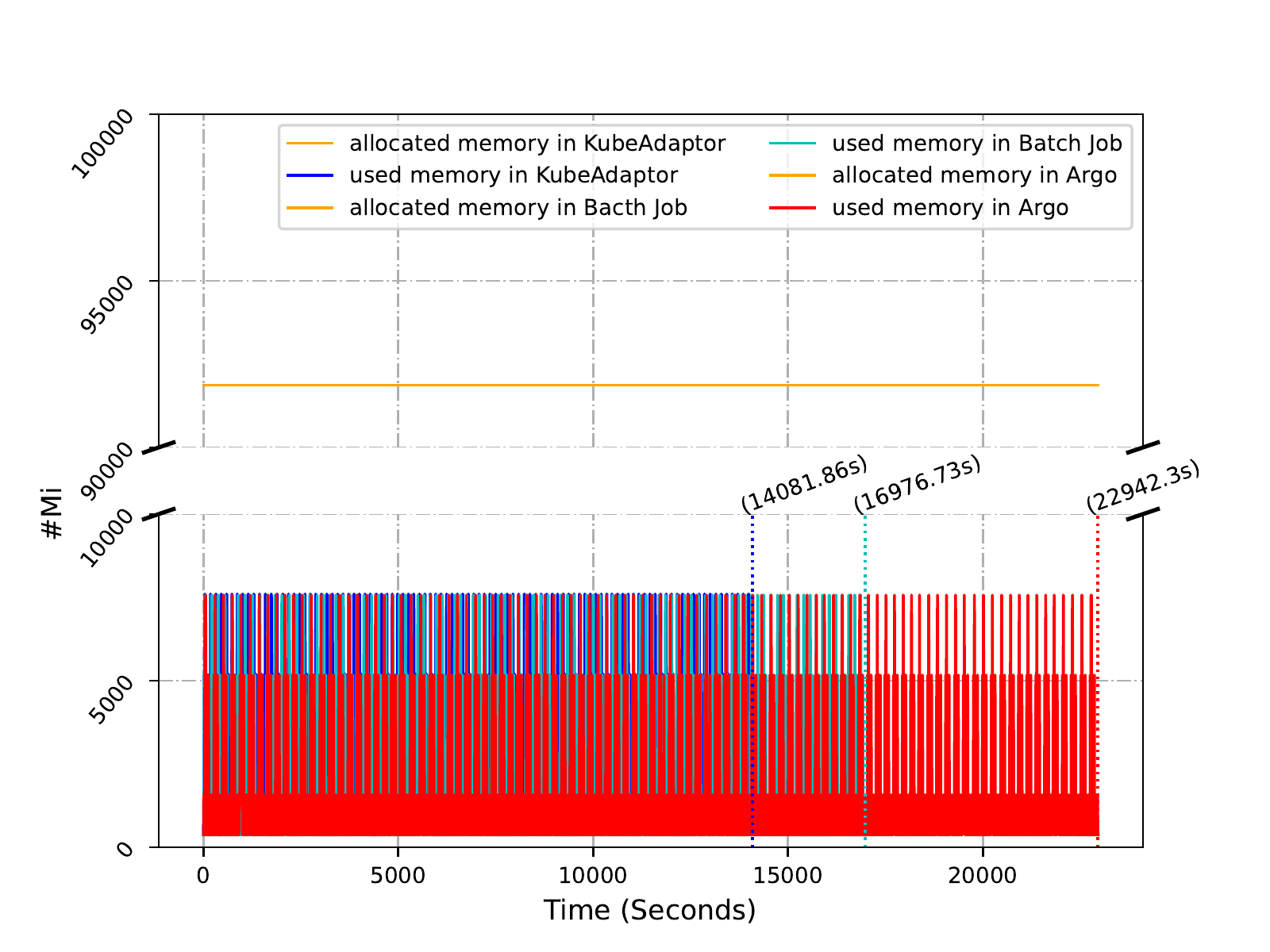}}
\hspace{0.01\linewidth}
\subfigure[Epigenomics]{\label{fig:mem25000:b}
\includegraphics[width=0.45\linewidth]{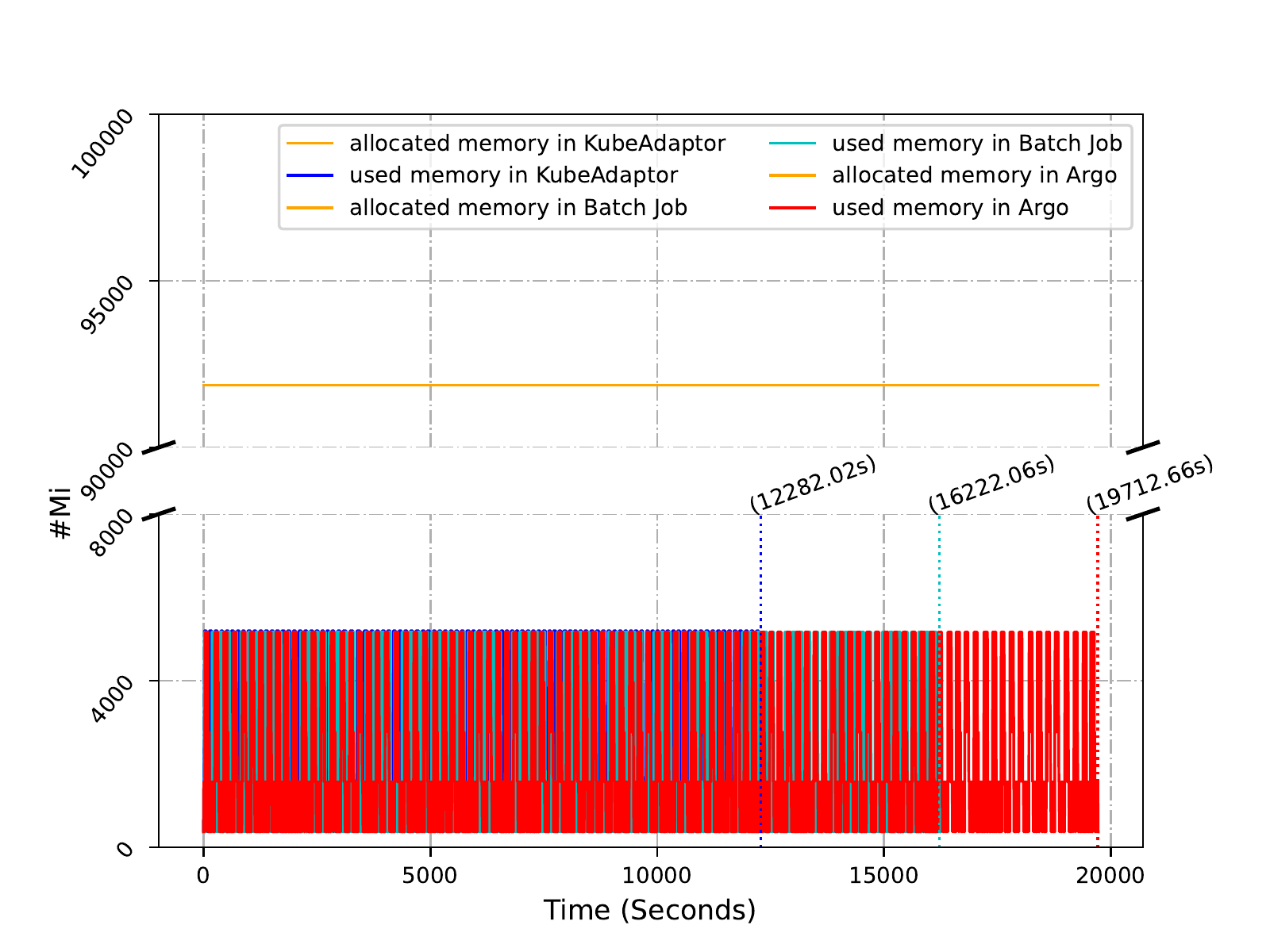}}
\subfigure[CyberShake]{\label{fig:mem25000:c}
\includegraphics[width=0.45\linewidth]{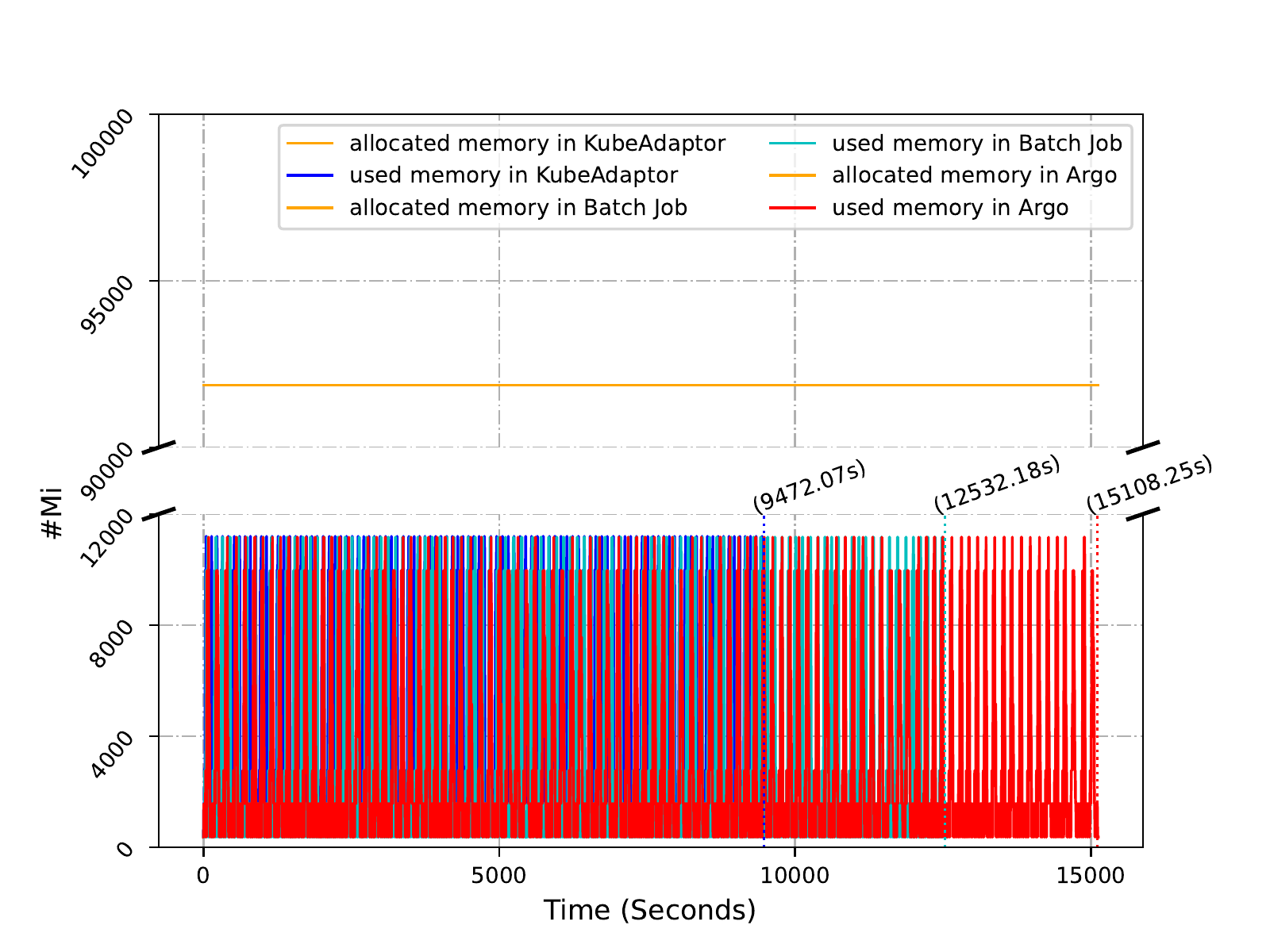}}
\hspace{0.01\linewidth}
\subfigure[LIGO]{\label{fig:mem25000:d}
\includegraphics[width=0.45\linewidth]{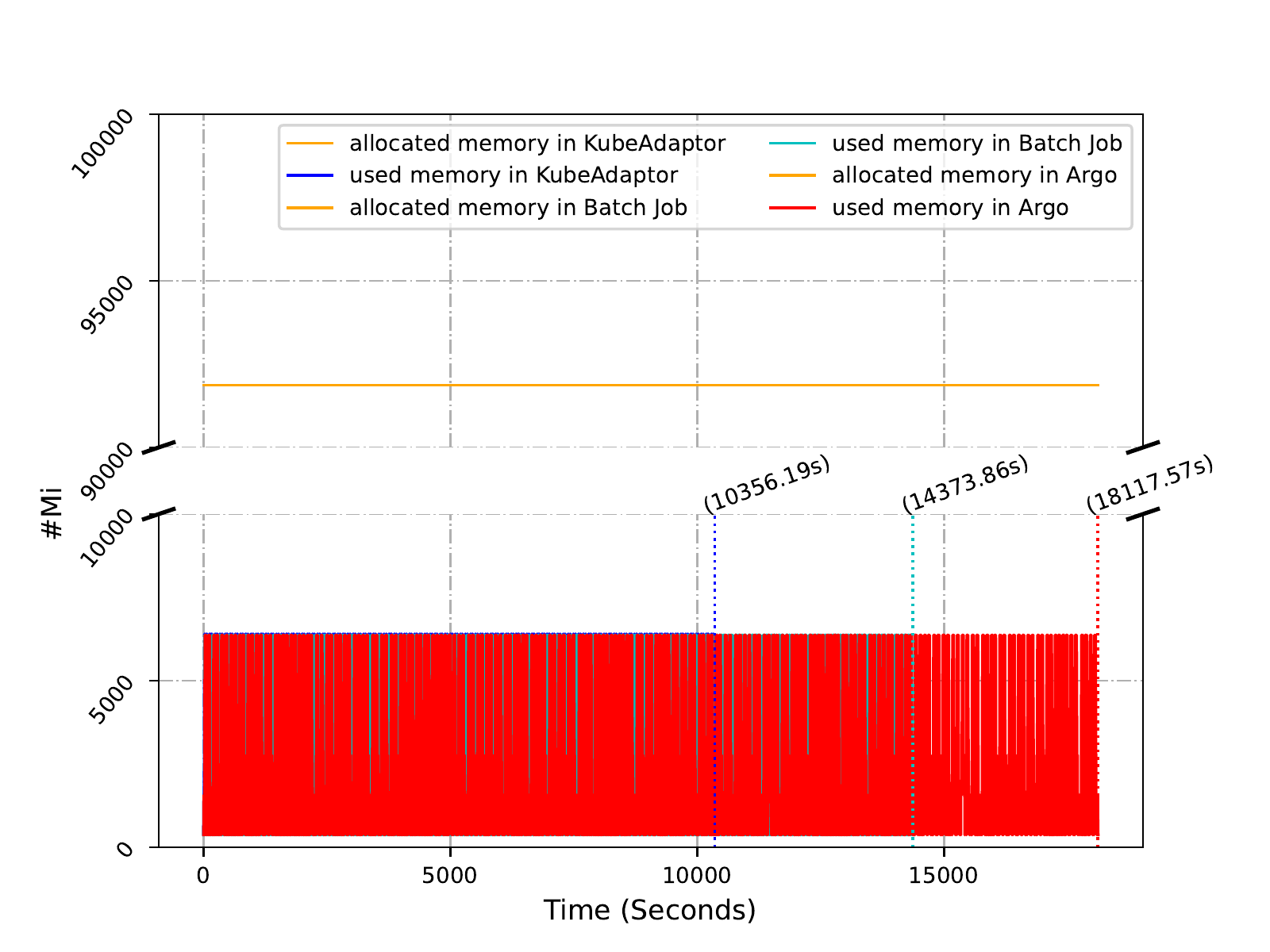}}
\caption{
  The memory usage rate of four real-world workflows in $100$ consecutive experiments. 
  The four subfigures show the curve of memory resource usage of four real-world workflows under 
three workflow submission approaches, respectively.   
    Except for the Master node, the other nodes participate in resource load and have $91872Mi$ 
    of memory in all. The orange curve in each subfigure depicts the number of allocatable memory in K8s cluster. 
    The blue, cyan, and red curves depict the number of memory used in KubeAdaptor, Batch Job, and Argo, 
respectively.
}
\label{fig:mem25000-usage}
\end{figure*}

  Under three distinct workflow submission approaches, Figure.~\ref{fig:averageWfTime} shows the workflow lifecycle 
  fluctuations and average workflow lifecycle of the four real-world workflows, respectively.
  We can observe that the KubeAdaptor is better than the other two approaches. 
  As expected, the Batch Job approach can not start the next batch 
  until the current task batch is fully completed. 
  This approach ignores the fact that some tasks in the next task batch 
  have met the execution conditions in advance, which prolongs the workflow lifecycle. 
  In addition, continuous deployment and cleanup of workflow tasks via {\itshape Kubectl} command require continual 
  checking of the status of the task pod, which increases the number of interactions with K8s apiserver and lengthens 
  the workflow lifecycle to some extent.
  Due to the unique internal logic processing of the Argo, the average workflow lifecycle of the Argo is higher 
  than that of the other two approaches. 
  Similar to Batch Job, continuous deployment and cleanup of workflows via the Argo binary tool also require continual 
  checking of the status of the task pod, which also increases the number of interactions with K8s apiserver and 
  prolongs the workflow life cycle to some extent.
  Understandably, the overall execution time of $100$ consecutive workflows captured by Batch Job and Argo 
 in this way is slightly larger than the overall execution time of $100$ consecutive workflows obtained 
 by watching resource changes shown in Figure~\ref{fig:cpu25000-usage} and Figure \ref{fig:mem25000-usage}.  
 It is due to the overall workflow execution time obtained by watching resource changes does not include 
 the time of the first workflow deployment and the last workflow cleanup. 
  Compared to the Batch Job and Argo, KubeAdaptor always obtains superior performance in terms of average 
  workflow lifecycle despite including the creation time and the binding time of PVC, 
  which benefits from the unique container construction function and event trigger mechanism 
  within the KubeAdaptor.
  Compared to the baseline Argo, the KubeAdaptor reduces the average workflow lifecycle 
  by $43.44\%$~(Montage), $43.65\%$~(Epigenomics), $44.86\%$~(CyberShake) and $48.98\%$~(LIGO) for the four 
  real-world workflows, respectively.

\paragraph{\bfseries{Resource Usage Comparison.}}
In this subsection, we develop a separate resource gathering module for K8s. 
This module aims to capture the state changes of underlying resources in the K8s cluster under each workflow 
submission approach and present the superior resource utilization characteristics of the KubeAdaptor 
through resource changes. 
This resource gathering module needs to be containerized and deployed on the K8s cluster in advance. 
We present the resource usage of each real-world workflow in consecutive $100$ experiments under three 
submission approaches and analyze the resource changes under each submission approach in the first $250$ seconds. 

\begin{figure*}
\centering
\subfigure[Montage]{\label{fig:cpu250:a}
\includegraphics[width=0.45\linewidth]{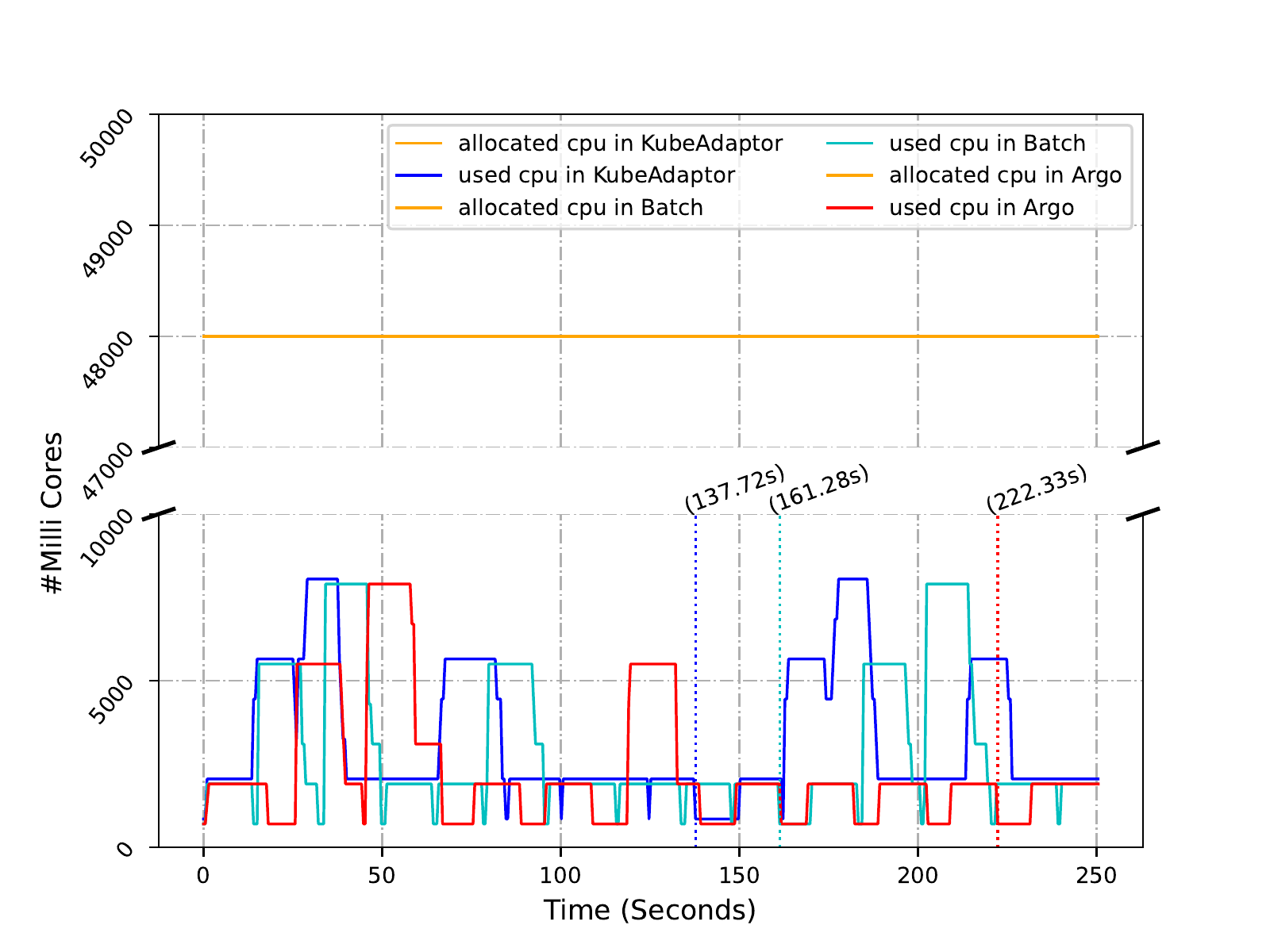}}
\hspace{0.01\linewidth}
\subfigure[Epigenomics]{\label{fig:cpu250:b}
\includegraphics[width=0.45\linewidth]{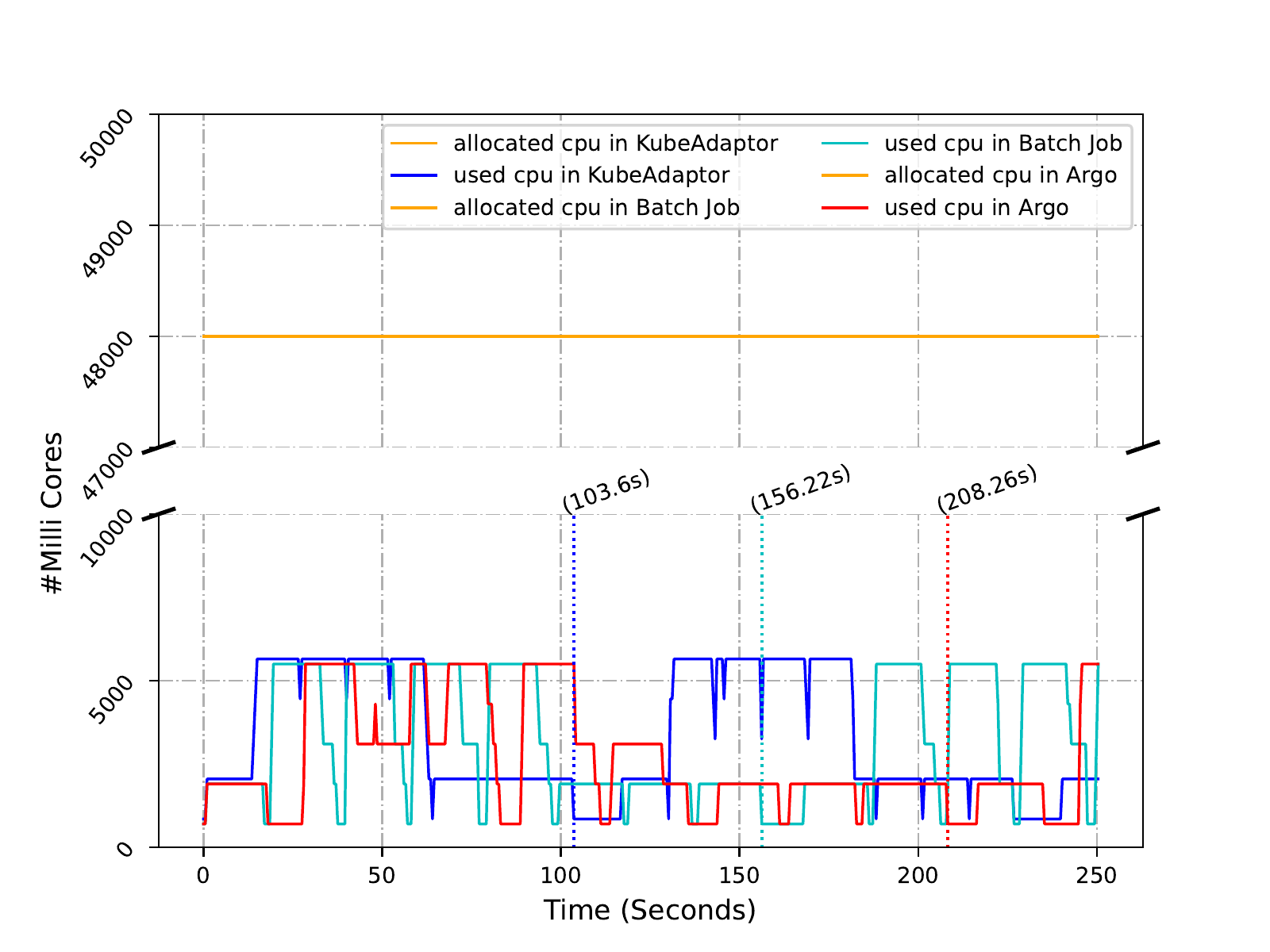}}
\subfigure[CyberShake]{\label{fig:cpu250:c}
\includegraphics[width=0.45\linewidth]{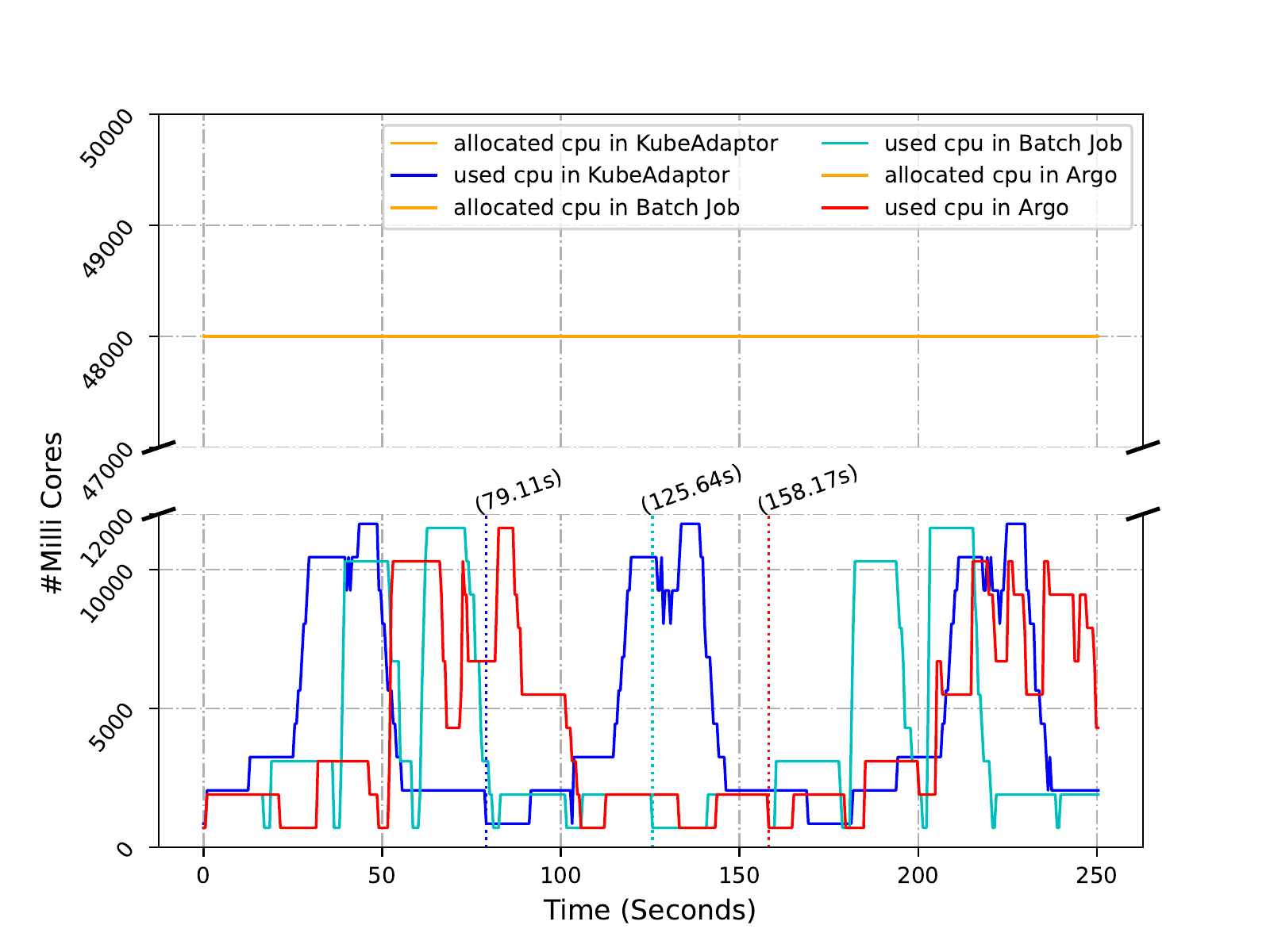}}
\hspace{0.01\linewidth}
\subfigure[LIGO]{\label{fig:cpu250:d}
\includegraphics[width=0.45\linewidth]{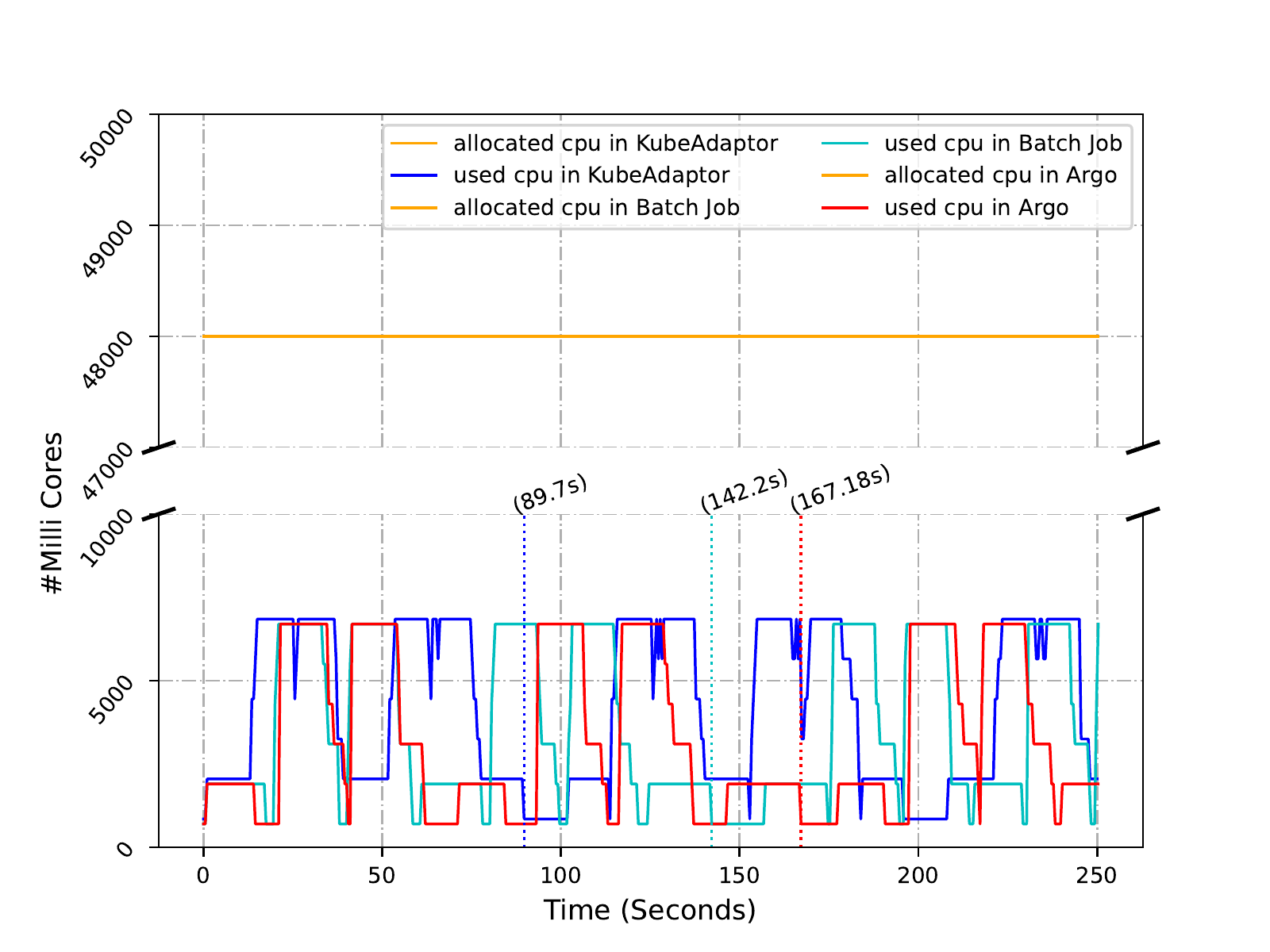}}
\caption{
The CPU usage rate of four real-world workflows in the first $250$ seconds. 
}
\label{fig:cpu250-usage}
\end{figure*}

\begin{figure*}
\centering
\subfigure[Montage]{\label{fig:mem250:a}
\includegraphics[width=0.45\linewidth]{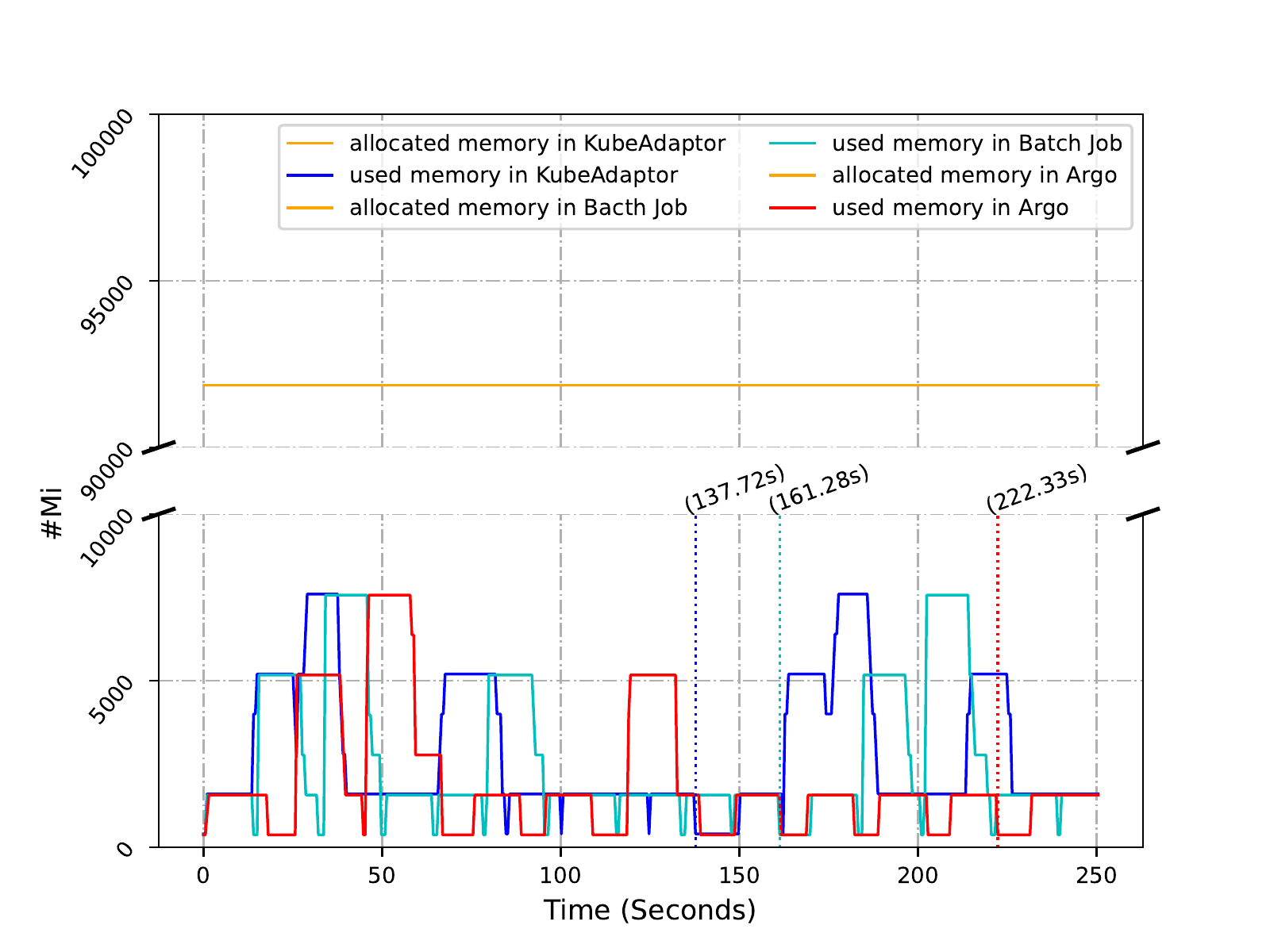}}
\hspace{0.01\linewidth}
\subfigure[Epigenomics]{\label{fig:mem250:b}
\includegraphics[width=0.45\linewidth]{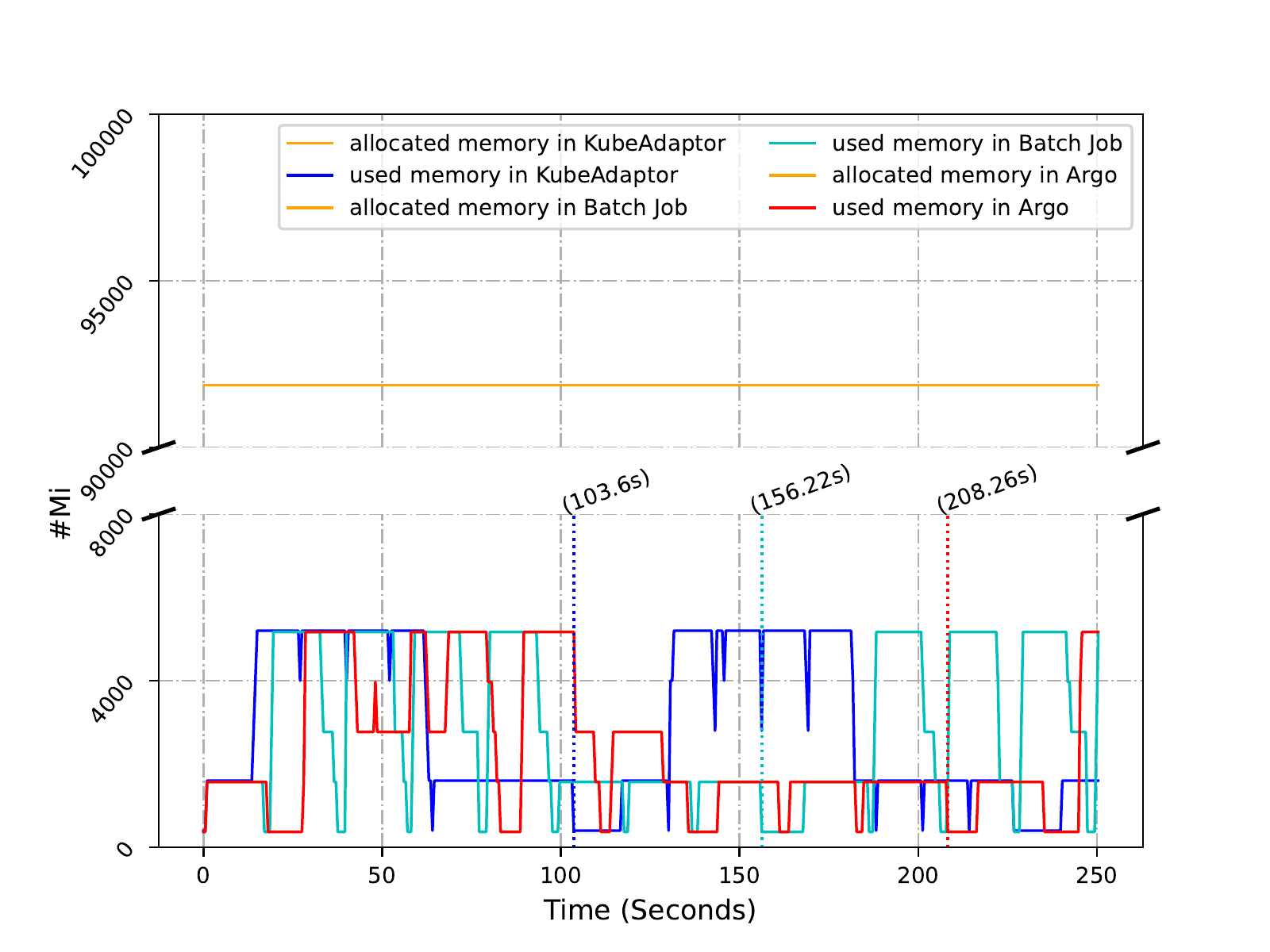}}
\subfigure[CyberShake]{\label{fig:mem250:c}
\includegraphics[width=0.45\linewidth]{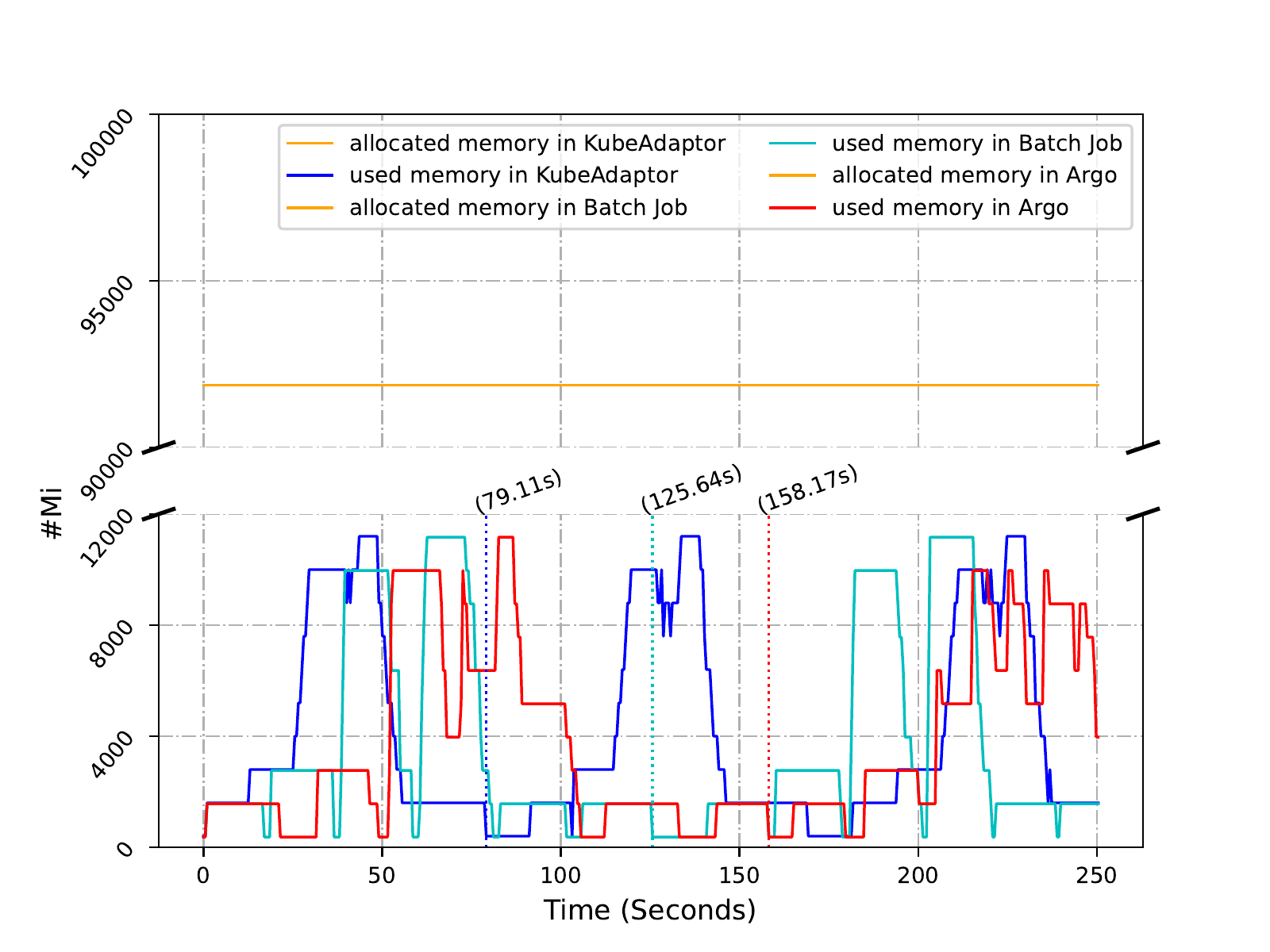}}
\hspace{0.01\linewidth}
\subfigure[LIGO]{\label{fig:mem250:d}
\includegraphics[width=0.45\linewidth]{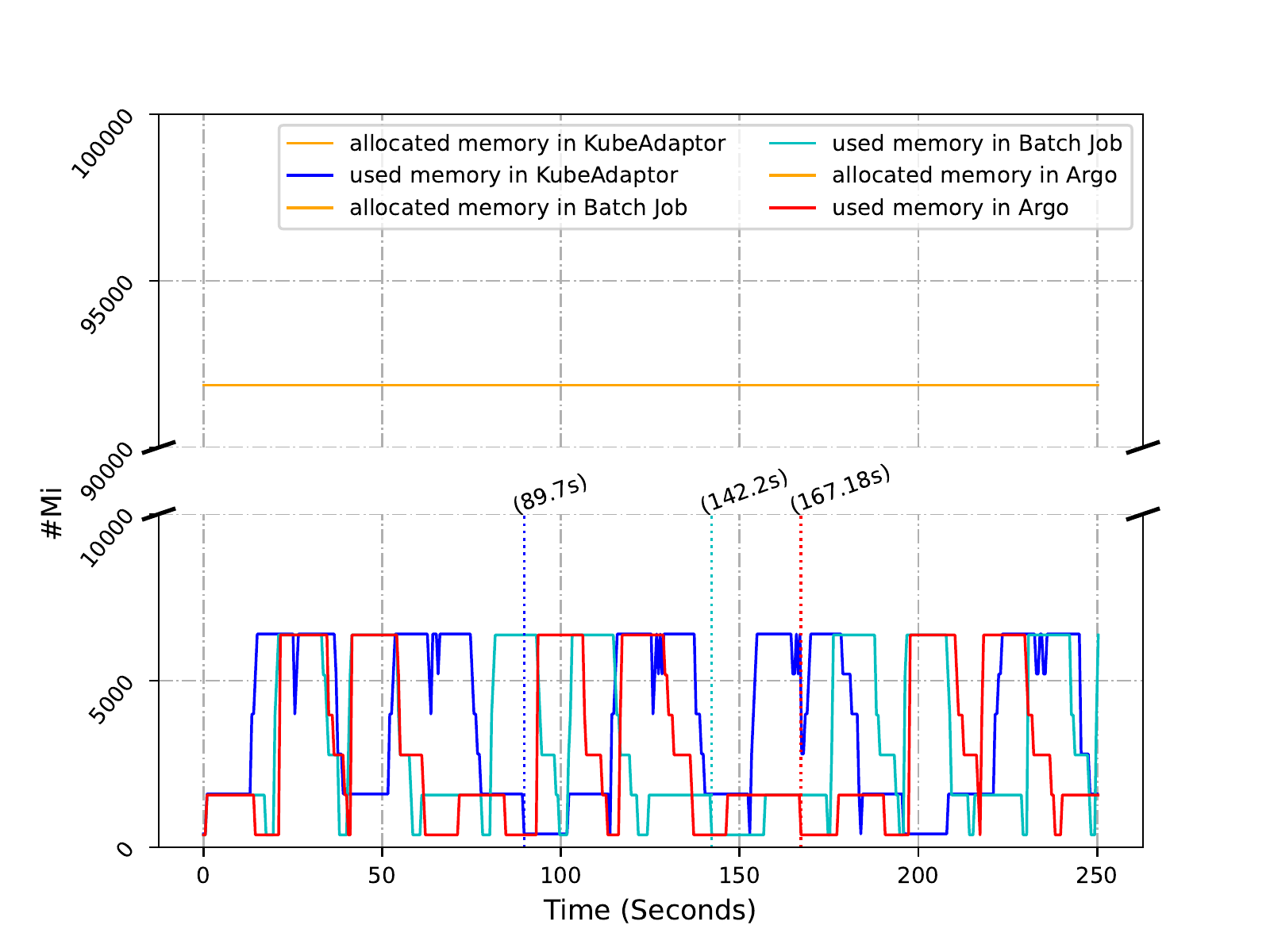}}
\caption{
  The memory usage rate of four real-world workflows in the first $250$ seconds.
}
\label{fig:mem250-usage}
\end{figure*}

The eight subfigures in Figure.~\ref{fig:cpu25000-usage} and Figure.~\ref{fig:mem25000-usage} 
show the resource usage curve of CPU and memory corresponding to four real-world 
workflows under three workflow submission approaches, respectively. 
For ease of comparison of three workflow submission approaches, the K8s cluster only accepts the experimental 
workflow load imposed by us.
By continuously executing each workflow 100 times under each submission method, we can obtain the overall consumed 
time of each workflow under each submission method, such as Montage ($14081.86s$ of KubeAdaptor, $16976.73s$ of Batch Job, and 
$22942.3s$ of Argo), Epigenomics ($12282.02s$ of KubeAdaptor, $16222.06s$ of Batch Job, and $19712.66s$ of Argo), 
CyberShake($9472.07s$ of KubeAdaptor, $12532.18s$ of Batch Job, and $15108.25s$ of Argo), and LIGO ($10356.19s$ of 
KubeAdaptor, $14373.86s$ of Batch Job, and $18117.57s$ of Argo). 
KubeAdaptor takes the least time, and Argo takes the most. 
The reason for this is consistent with the average workflow lifecycle shown in Figure~\ref{fig:averageWfTime}. 
In terms of CPU resource usage, four workflows with concurrent tasks can consume up to $8050m$, $5650m$, $11650m$,
and $6850m$, respectively. 
For memory resource usage, four workflows with concurrent tasks can consume up to $7600Mi$, $5200Mi$, $11200Mi$, 
and $6400Mi$, respectively.
In Figure.~\ref{fig:cpu250-usage} and Figure.~\ref{fig:mem250-usage}, we will zoom in the resource usage of the first $250s$ 
for each workflow in Figure.~\ref{fig:cpu25000-usage} and Figure.~\ref{fig:mem25000-usage} under three workflow submission methods. 

Figure.~\ref{fig:cpu250-usage} and Figure.~\ref{fig:mem250-usage} show the usage of CPU and memory of four real-world 
workflows under three workflow submission approaches in the first $250$ seconds, respectively. 
Four subfigures in Figure.~\ref{fig:cpu250-usage} correspond to the four subfigures in Figure.~\ref{fig:cpu25000-usage} 
respectively, and the four subfigures in Figure.~\ref{fig:mem250-usage} correspond to the four subfigures 
in Figure.~\ref{fig:mem25000-usage} respectively. 
Across the complete timeline of CPU usage and memory usage in Figure.~\ref{fig:cpu250-usage} and 
Figure.~\ref{fig:mem250-usage}, the resource usage curve of each workflow submission method reflects the 
workflow lifecycle in the first $250s$, and both are consistent in curve shape.
For each subfigure in Figure.~\ref{fig:cpu250-usage} and Figure.~\ref{fig:mem250-usage}, the resource usage curves 
in task scheduling of a workflow under three workflow submission approaches are consistent in the curve shape. 
And the differences lie in the switching time between tasks in each workflow, the average execution time of 
the task pod, and workflow lifecycle.
In the K8s cluster, the allocatable CPU is $48000m$, and the allocatable memory is $91872Mi$. 
In each subfigure of Figure.~\ref{fig:cpu250-usage} and Figure.~\ref{fig:mem250-usage}, each protruding time of the 
curve represents the execution time of task pod. 
In addition, the \verb|requests| field of task pod includes $1200m$ CPU and $1200Mi$ memory, and the 
\verb|limits| field have the same parameter as the \verb|requests| field. 

We focus only on the resource usage changes of each workflow during the first workflow lifecycle under 
the corresponding workflow submission method. 
From Figure.~\ref{fig:cpu250-usage} and Figure.~\ref{fig:mem250-usage}, we can obtain the first 
workflow lifecycle of each workflow under each submission method, such as Montage ($137.72s$ of KubeAdaptor, 
$161.28s$ of Batch Job, and $222.33s$ of Argo), Epigenomics ($103.6s$ of KubeAdaptor, $156.22s$ of Batch Job, 
and $208.26s$ of Argo), CyberShake ($79.11s$ of KubeAdaptor, $125.64s$ of Batch Job, and $158.17s$ of Argo), 
and LIGO ($89.7s$ of KubeAdaptor, $142.2s$ of Batch Job, and $167.18s$ of Argo). 
Across the first workflow lifecycle of each workflow under each workflow submission approach, we can observe 
that the switching process of the task pod in KubeAdaptor is too fast to capture the resource changes within 
the sampling period of $0.5s$. 
 Nevertheless, the switching process of the task pod in Batch Job and Argo is relatively long. 
Compared to Batch Job and Argo, the KubeAdaptor has a shorter switching time of task pod. 
 It is due to that the event trigger mechanism makes KubeAdaptor respond to the state changes of underlying 
 resources of K8s in real-time and rapidly invoke the creation or destruction of the task pod. 
For the performance of switching time of task pod, batch submission and deletion modes via Kubectl command 
line in Batch Job, and unique triggering logic of task pod in Argo, consume too much time, far inferior to 
the event trigger mechanism in KubeAdaptor.

\begin{figure}[h]
  \centering
  \includegraphics[width=\linewidth]{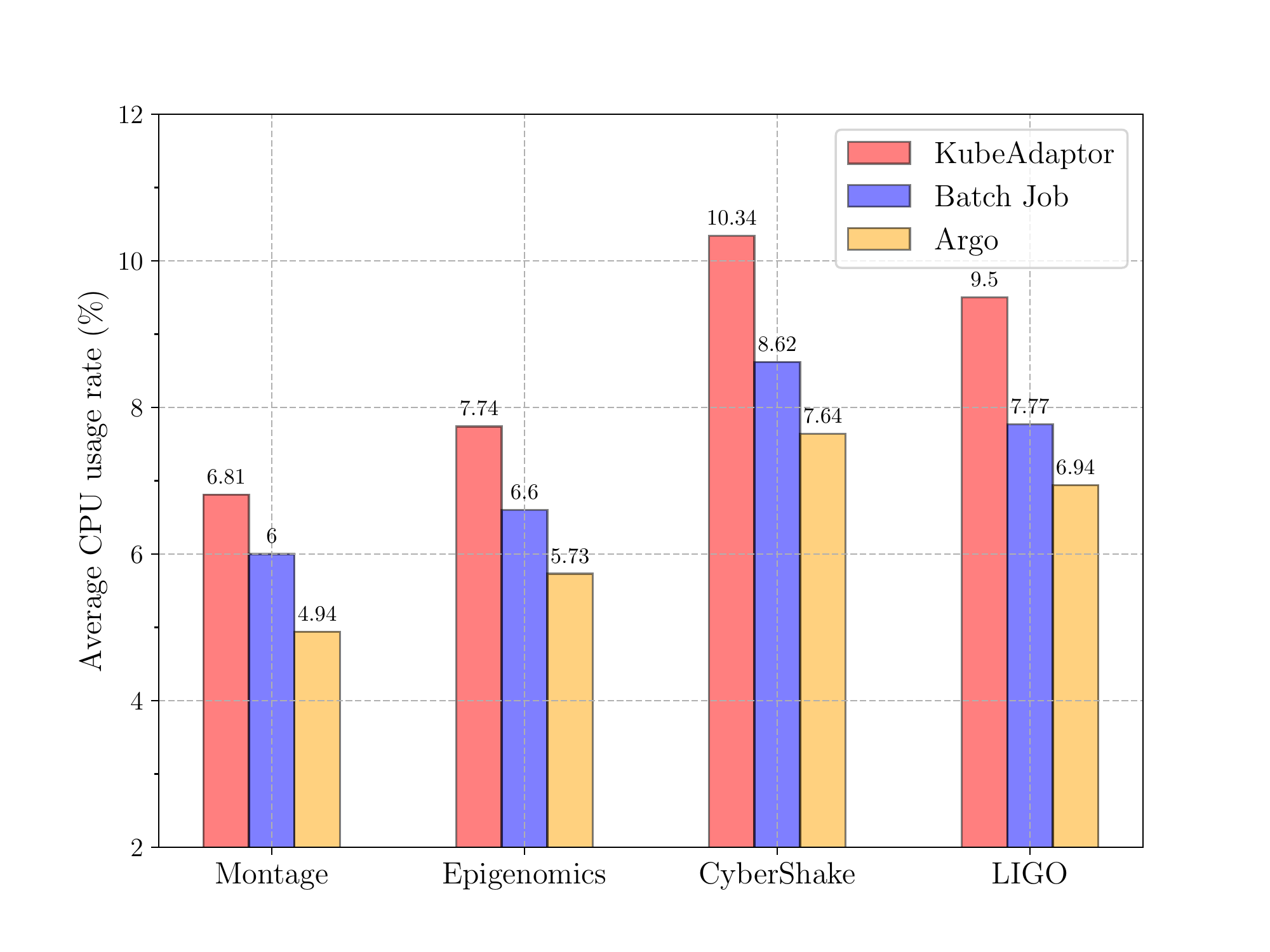}
  \caption{Average CPU usage rate. }
  \label{fig:aveargeCpuRate}
\end{figure}
\begin{figure}[h]
  \centering
  \includegraphics[width=\linewidth]{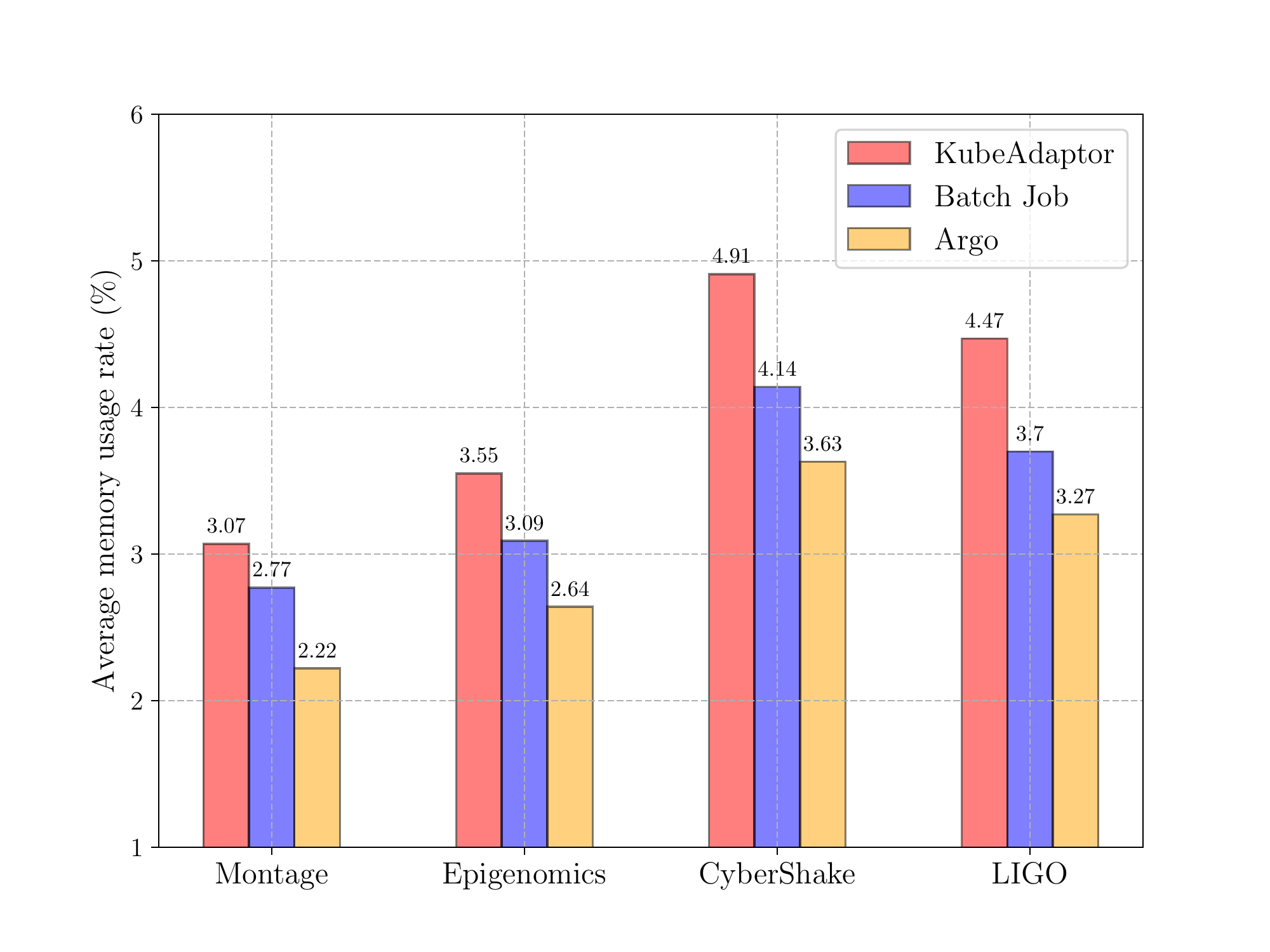}
  \caption{Average memory usage rate. }
  \label{fig:aveargeMemRate}
\end{figure}

According to resource usage data of three workflow submission methods in each subfigure of 
Figure.~\ref{fig:cpu250-usage} and Figure. \ref{fig:mem250-usage}, across the first lifecycle of workflow, 
we can obtain average CPU usage rate and average memory usage rate shown in Figure.~\ref{fig:aveargeCpuRate} 
and Figure.~\ref{fig:aveargeMemRate}. 
In the average CPU usage rate and memory usage rate of the first workflow lifecycle, we can observe that 
the KubeAdaptor is the best, followed by Batch Job, and Argo is the worst.  
KubeAdaptor obtains a shorter switching time of the task pod through the event trigger mechanism.
 The shorter the switching time of the task pod, the higher the resource usage.  As the number of workflow 
 tasks increases, the shorter switching time of the task pod ensures the KubeAdaptor maintains an advantage 
 in resource usage.

\section{Conclusion and Future Work}	
In this paper, our proposed KubeAdaptor for K8s implements workflow containerization following the optimized 
task order of workflow scheduling algorithms and integrates workflow systems with the K8s.
The construction and running of task pods in workflow depend on the Informer-based resource object monitoring 
scheme and the event triggering mechanism.
Benefiting from the event triggering mechanism and its superior logic design, KubeAdaptor ensures the rapid response to the underlying 
 resource state changes of K8s and restricts the out-of-order scheduling, scattered scheduling, and 
 unpredictability of native K8s scheduler.
On the aspects of average workflow lifecycle, average task pod execution time, and resource usage rate, 
our customized KubeAdaptor outperforms the Batch Job and Argo.

The work presented in this paper, KubeAdaptor, as a logic interface for docking K8s, provides a practical 
framework for the engineering practice of containerized workflow systems on K8s. 
In the future, we will investigate the case study of cloud workflow scheduling with KubeAdaptor enabled in 
cloud control systems, develop a workflow description tool compatible with KubeAdaptor, and further implement 
a graphical user interface to input workflow through drag and drop widgets. 
In addition, the KubeAdaptor's core functions, event triggering mechanism, and gRPC communication mechanism 
also provide a practical solution for cloud-edge task migration under the cloud-edge cooperation framework.

\section{Acknowledgments}
This work is supported by the National Key Research and Development Program of China (Grant No. 2018YFB1003700).










\bibliographystyle{elsarticle-num-names}
\bibliography{cas-refs}


\begin{wrapfigure}{l}{25mm} 
  \includegraphics[width=1in,height=1.25in,clip,keepaspectratio]{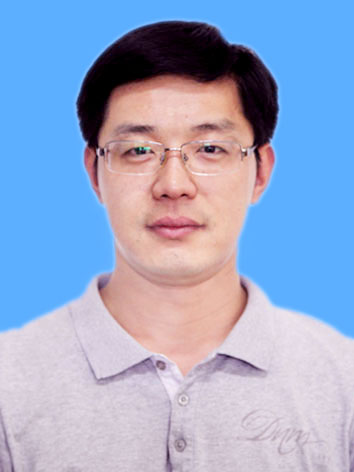}
\end{wrapfigure}\par
\textbf{Chenggang Shan}~received the M.S. degree in computer applied technology from Qiqihr University, China, in 2007. 
He is working toward the Ph.D. degree with the School of Automation, Beijing Institute of Technology, Beijing, 
China. He was an associate professor with the School of Artificial Intelligence, Zaozhuang University, China, 
in 2017. His research interests include networked control systems, cloud computing, cloud-edge collaboration, 
wireless networks. He is a member of Chinese Computer Federation (CCF).\par
\newpage
\begin{wrapfigure}{l}{25mm} 
  \includegraphics[width=1in,height=1.25in,clip,keepaspectratio]{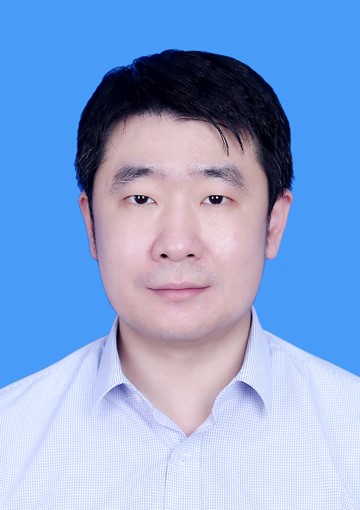}
   \end{wrapfigure}\par
\textbf{Guan Wang}   
received the M.S. degree in CTS from University of Jinan, China, in 2014. He currently working toward the Ph.D. 
in the School of Automation, Beijing Institute of Technology, China. He was a lecturer in the School of 
Information Science and Engineering, University of Zaozhuang, China in 2019. 
His research interests lie in the areas of networking systems, cloud computing, gene expression data, 
and machine learning.\par

\begin{wrapfigure}{l}{25mm} 
  \includegraphics[width=1in,height=1.25in,clip,keepaspectratio]{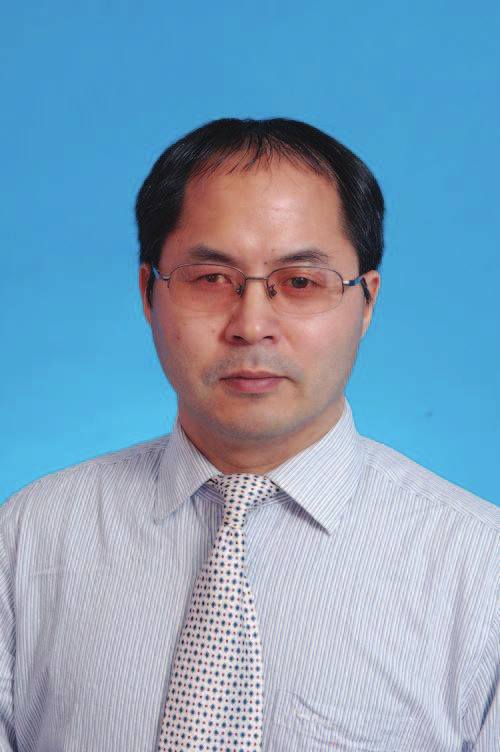}
  \end{wrapfigure}\par
\textbf{Yuanqing Xia}  
(M'15-SM'16) received the Ph.D. degree in Control Theory and Control Engineering from Beijing University of 
Aeronautics and Astronautics, Beijing, China, in 2001.
From November 2003 to February 2004, he was with the National University of Singapore as a Research Fellow, 
where he worked on variable structure control. Since 2004, he has been with the School of Automation, 
Beijing Institute of Technology, Beijing, first as an Associate Professor, then, since 2008, as a Professor. 
His current research interests are in the fields of networked control systems, robust control and signal 
processing, active disturbance rejection control.
\par

\begin{wrapfigure}{l}{25mm} 
  \includegraphics[width=1in,height=1.25in,clip,keepaspectratio]{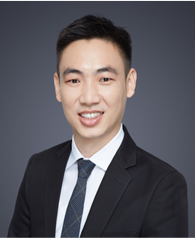}
  \end{wrapfigure}\par
\textbf{Yufeng Zhan}
received his Ph.D. degree from Beijing Institute of Technology (BIT), Beijing,
China, in 2018. He is currently an assistant professor in the School of Automation with BIT.
Prior to join BIT, he was a post-doctoral fellow in the Department of Computing with The Hong
Kong Polytechnic University. 
His research interests include networking systems, game theory, and machine learning.\par

\begin{wrapfigure}{l}{25mm} 
  \includegraphics[width=1in,height=1.25in,clip,keepaspectratio]{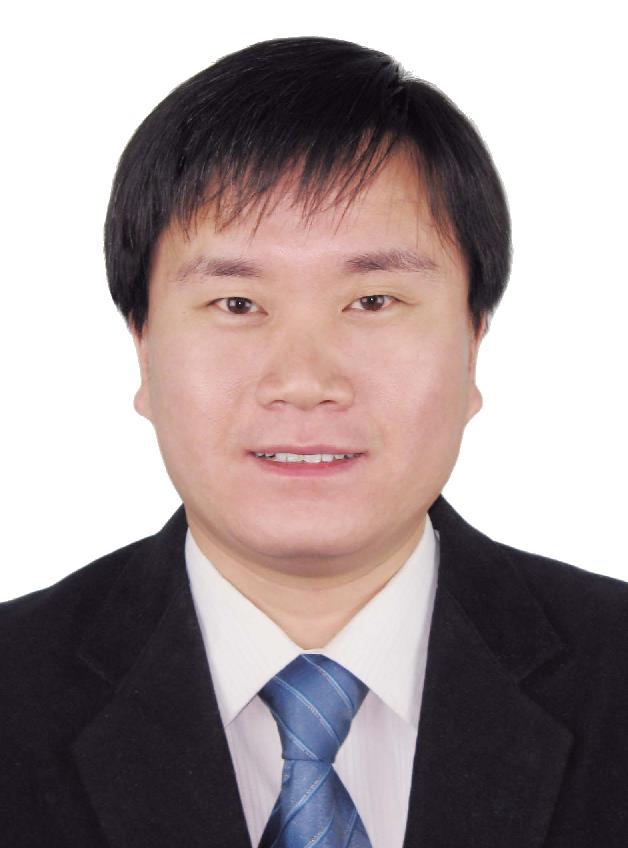}
       \end{wrapfigure}\par
\textbf{Jinhui Zhang} 
received the Ph.D. degree in Control Science and Engineering from Beijing Institute of Technology, Beijing, 
China, in 2011. He was a Visiting Fellow with the School of Computing, Engineering \& Mathematics, 
University of Western Sydney, Sydney, Australia, from February 2013 to May 2013. 
He was an Associate Professor in the Beijing University of Chemical Technology, Beijing, 
from March 2011 to March 2016, a Professor in the School of electrical and automation engineering, 
Tianjin University, Tianjin, from April 2016 to September 2016. He joined Beijing Institute of Technology 
n October 2016, where he is currently an Tenured Professor. His research interests include networked control 
systems and composite disturbance rejection control.\par


\end{document}